\documentclass{ametsocV6.1}

\title{Resonant excitation of Kelvin waves by interactions of subtropical Rossby waves and the zonal mean flow}

\newtheorem{proposition}{Proposition}
\authors{Katharina M. Holube,\aff{a}\correspondingauthor{Katharina M. Holube, katharina.holube@uni-hamburg.de}
Frank Lunkeit,\aff{a}
Sergiy Vasylkevych,\aff{a}
Nedjeljka \v{Z}agar\aff{a}}

\affiliation{\aff{a}{Meteorological Institute, Center for Earth System Research and Sustainability, Universit\"at Hamburg, Hamburg, Germany}}

\abstract{Equatorial Kelvin waves can be affected by subtropical Rossby wave dynamics. Previous research has demonstrated the Kelvin wave growth in response to subtropical forcing and the resonant growth due to eddy momentum flux convergence. However, the relative importance of the wave-mean flow and wave-wave interactions for the Kelvin wave growth compared to the direct wave excitation by the external forcing has not been made clear. This study demonstrates the resonant Kelvin wave excitation by interactions of subtropical Rossby waves and the mean flow using a spherical shallow-water model. The use of Hough harmonics as basis functions makes Rossby and Kelvin waves prognostic variables of the model and allows the quantification of terms contributing to their tendencies in physical and wave space.\\
The simulations show that Kelvin waves are resonantly excited by interactions of Rossby waves and the balanced zonal mean flow in the subtropics, provided the Rossby and Kelvin wave frequencies, which are modified by the mean flow, match. The resonance mechanism is substantiated by analytical expressions. The Kelvin wave tendencies are caused by velocity and depth tendencies: The velocity tendencies due to the meridional advection of zonal mean velocity can be outweighed by the zonal advection of Rossby wave velocity or by the depth tendencies due to Rossby wave divergence. Identifying the resonant excitation mechanism in data should contribute to the quantification of Kelvin wave variability originating in the subtropics.}

\begin{document}

\maketitle

\statement
This study seeks to understand how Kelvin waves, which are eastward-propagating disturbances in the tropical atmosphere, are connected to Rossby wave dynamics in the subtropics. Using idealized simulations, the Kelvin wave excitation is explained as a resonance effect due to interactions of Rossby waves and the zonal mean flow. The mechanism contributes to the understanding of atmospheric wave interactions and extratropical effects on the tropics. Further work searching for evidence of the new mechanism in atmospheric data may shed a new light on subseasonal variability in the tropics, such as the Madden-Julian Oscillation. 

%
%
%
%
%
%


\section{Introduction}
Many aspects of the large-scale atmospheric circulation are explained by reduced models which target particular regimes and scales of interest. These reduced models are usually derived separately for the extratropics, i.e.~the quasi-geostrophic regime \citep[e.g.][]{Dolaptchiev2013}, and the tropics \citep[e.g.][]{Majda2003}. A reduced model suitable for studying subtropical processes, which involve both Rossby waves and equatorial waves, is a model based on the nonlinear shallow-water equations on the sphere \citep[e.g.][]{Cho_Polvani_1996,KitamuraIshioka2007,Kraucunas.Hartmann_2007,barpanda_role_2023}. Such a model offers insights into wave-wave and wave-mean flow interactions that are more challenging to obtain from tropical or mid-latitude $\beta$-plane models. In the present study, a spherical shallow-water model is used to investigate the excitation of the equatorial Kelvin wave (KW) by subtropical Rossby wave dynamics.

The extratropical influence on tropical processes has been addressed in a number of studies since \citet{Webster.Holton_1982} established that the influence can take place through equatorward-propagating Rossby waves in the region of westerlies at the equator (i.e.~the westerly duct). Early studies found that the amplitude and structure of laterally forced equatorial waves strongly depend upon the tropical mean zonal wind \citep{Zhang.Webster_1992,Zhang_1993}. In the studies by \citet{Zhang.Webster_1992} and \citet{Zhang_1993} KWs could have been be excited by wave-mean flow interactions as well as by direct forcing because the truncated version of their linearized shallow-water model did not permit meridional geopotential advection and wave-wave interactions. They also found that the extratropical influence on the KWs was stronger in equatorial easterlies than in westerlies. \citet{barpanda_role_2023} employed a spherical shallow-water model with topography along the equator to study the influence of the subtropical jet on the coupling between KWs and subtropical Rossby waves. By diagnosing terms of the linearized vorticity equation in the steady state, these authors showed that KWs are affected by the advection of background absolute vorticity by the Rossby wave meridional velocity.
In primitive-equation model simulations conducted by \citet{Hoskins.Yang_2000}, KWs were excited by eastward-propagating subtropical heating or vorticity forcing. The Kelvin wave response was strongest when the forcing frequency was close to the frequency of the KWs, i.e.~when the KW was nearly in resonance with the forcing. However, the wave could have been excited by wave-mean flow interactions, by wave-wave interactions, and directly by the external forcing.

The present study elucidates the role of subtropical wave-mean flow and wave-wave interactions on the KW excitation by idealized numerical simulations without direct effects of the external forcing on the Kelvin wave. We show that the Kelvin wave is excited by interactions of Rossby waves and the zonal mean flow in the subtropics when the phase speed of the eastward-shifted Rossby waves matches the KW phase speed.

Our analysis of the KW growth due to resonance is facilitated by a novel numerical model that solves the spherical shallow-water equations using the Hough harmonics as basis functions \citep{Vasylkevych.Zagar_2021}. The Hough harmonics are  eigensolutions of the spherical shallow-water equations linearized around the state of rest. In this formulation, the KW is defined as the part of the simulated circulation that projects on the slowest eastward-propagating 
component of the basis function set. For small mean fluid depths, the spherical eigensolution for the KW is almost identical to the Kelvin wave solution on the equatorial $\beta$ plane, i.e.~\citet{Matsuno_1966}'s Kelvin wave \citep[e.g.][]{Boyd2018,Zagar.etal_2022}. On the sphere, KWs have a small meridional velocity component, are weakly dispersive and their trapping is defined by both the mean (or equivalent) depth and the zonal wavenumber \citep{Boyd.Zhou_2008}.

The nearly non-dispersive nature of the large-scale Kelvin waves facilitates their identification in observations and weather and climate models, especially in relation to deep tropical convection \citep[e.g.][]{Kiladisetal2009,Knippertz.etal_2022}.
Filtering Kelvin waves associated with subtropical processes is more complex in part because the amplitude of quasi-geostrophic wave dynamics well exceeds that of equatorial wave perturbations. Nevertheless, a significant enhancement of KW activity in parts of the tropics was found to be coupled with stronger eastward-propagating Rossby wave activity in the subtropics \citep{Straub.Kiladis_2003,Tulich.Kiladis_2021,Cheng.etal_2022}.
When subtropical Rossby waves break at their critical line, they can affect KWs through the convergence of eddy momentum fluxes \citep{Tulich.Kiladis_2021}. The subtropical jet advects the Rossby waves eastward, aligning their phase speed with that of the KWs. The phase of the eddy momentum flux convergence thus varies little relative to the propagating KWs, so that they grow through a resonance effect \citep{Cheng.etal_2022}.  During austral winter, the extratropical effects can exceed the influence of local moist thermodynamics on KWs over the tropical Pacific \citep{Straub.Kiladis_2003}.

The eddy momentum flux analyzed in reanalysis data in the above mentioned studies belongs to the wave-wave interaction process. As a first step towards quantifying the role of this process compared to wave-mean flow interactions and direct forcing on the KW variability, we perform idealized experiments which do not contain KWs in the initial state and which allow the quantification of terms contributing to the KW growth. 
In what follows, we first present the Transient Inertia-Gravity And Rossby wave model \citep[TIGAR;][]{Vasylkevych.Zagar_2021} and the simulation setup (section \ref{sect:method}).
The linear Rossby wave response to the forcing is described in section~\ref{sect:linrossby}. The excitation of KWs as a resonance effect is explained in section~\ref{sect:results}. The conclusions and outlook are given in section~\ref{sect:conclusions}.

\section{Modeling setup}
\label{sect:method}
In the following, the key features of TIGAR and the analysis method of the wave-wave and wave-mean flow interactions are outlined. Subsequently, the zonal mean flow, the external forcing and the parameters of our simulation setup are described.
\subsection{Model formulation and energy equation}
\label{sect:model}
TIGAR solves the rotating shallow-water (RSW) equations in spherical coordinates $(\lambda, \varphi) \in [0,2\pi)\times(-\pi/2,\pi/2)$, which in non-dimensional form read
\begin{subequations}
\begin{eqnarray}
    \frac{\partial u}{\partial t} &+& \gamma \mathbf{V}\cdot \nabla u - \gamma u\upsilon\tan\varphi - \upsilon\sin\varphi +\frac{\gamma}{\cos\varphi}\frac{\partial h}{\partial\lambda} = F_u + Q_u\, ,\\
    \frac{\partial \upsilon}{\partial t} &+& \gamma \mathbf{V}\cdot \nabla \upsilon + \gamma u^2\tan\varphi + u\sin\varphi +\gamma\frac{\partial h}{\partial\varphi} = F_\upsilon + Q_\upsilon\, ,\\
    \frac{\partial h}{\partial t} &+& \gamma \mathbf{V}\cdot \nabla h +\gamma (h+1) \nabla\cdot \mathbf{V} = F_h + Q_h \,.
\end{eqnarray}
\label{eq:SWM_physical}
\end{subequations}
The horizontal velocity components $\mathbf{V} = (u,\upsilon)^\intercal$ and the fluid depth $h$ can be expressed as the state vector $\mathbf{X} = (u,\upsilon,h)^\intercal$. The non-dimensional variables are obtained as in \citet{Vasylkevych.Zagar_2021} by normalizing the depth deviation from the mean depth $D$ with $D$, whereas the velocities are normalized with $\sqrt{gD}$. The parameter $\gamma = \frac{\sqrt{gD}}{2a\Omega}$ contains the free parameters of the system: the Earth's radius $a$, gravity $g$ and rotation rate $\Omega$, and $D$, representing the equivalent depth of one vertical mode in the troposphere and stratosphere. The spectral viscosity $\mathbf{F} = (F_u, F_\upsilon, F_h)^\intercal$ damps the smallest resolved scales for all three variables, and it is defined following \citet{gelb_gleeson_2001}. The external forcing is denoted $\mathbf{Q} = (Q_u, Q_\upsilon, Q_h)^\intercal$.

TIGAR is a pseudospectral model that uses the Hough harmonics $\mathbf{H}_{n,l}^k(\lambda,\varphi)$ as basis functions, which are eigensolutions of Eq.~(\ref{eq:SWM_physical}), linearized around a state of rest with depth $D$ \citep{Longuet-Higgins_1968}. This linearization reduces Eq.~(\ref{eq:SWM_physical}) to 
\begin{equation} \label{LTE}
    \frac{\partial}{\partial t} \mathbf{X} + \mathbf{L}\mathbf{X} = 0,
\end{equation}
with 
\begin{equation}
    \mathbf{L} = \begin{pmatrix}
        0 & -\sin\varphi & \frac{\gamma}{\cos\varphi}\frac{\partial}{\partial \lambda}\\
        \sin\varphi & 0 & \gamma \frac{\partial}{\partial \varphi}\\
        \frac{\gamma}{\cos\varphi} \frac{\partial}{\partial\lambda} & \frac{\gamma}{\cos\varphi} \frac{\partial}{\partial\varphi}(\cos\varphi()) & 0
\end{pmatrix} \, . \label{eq:linear_operator}
\end{equation}
The Hough harmonics fulfill $\mathbf{L}\mathbf{H}_{n,l}^k = i\nu_{n,l}^k\mathbf{H}_{n,l}^k$ with eigenfrequency $\nu_{n,l}^k$. For each zonal wavenumber $k$, there is a range of meridional modes $n$ for three types of wave solutions: the eastward-propagating inertia-gravity (IG) modes ($l=1$), westward-propagating IG modes ($l=2$) and Rossby modes ($l=3$). The meridional structure of $\mathbf{H}_{n,l}^k(\lambda,\varphi) = \mathbf{\Theta}_{n,l}^k(\varphi)e^{ik\lambda}$ is defined by the Hough functions $\mathbf{\Theta}_{n,l}^k(\varphi)=\left(U_{n,l}^k(\varphi), iV_{n,l}^k(\varphi), Z_{n,l}^k(\varphi)\right)^\intercal$. The Kelvin mode $\mathbf{H}_{0,1}^k$ is the lowest meridional mode $(n=0)$ of the eastward-propagating inertia-gravity modes. See \citet{Swarztrauber.Kasahara_1985} and \citet{Kasahara2020} for details.

Expressing Eq.~(\ref{eq:SWM_physical}) in vector notation yields 
\begin{equation}
\frac{\partial}{\partial t} \mathbf{X} + \mathbf{L}\mathbf{X} = \mathbf{N} + \mathbf{Q}  +\mathbf{F}\, .
\label{eq:SWM}
\end{equation}
The term $\mathbf{N} = (N_u,N_\upsilon,N_h)^\intercal$ contains the interactions of different modes, as discussed in section \ref{sect:method}\ref{sect:wavemeanflow_wavewave}.

The expansion of Eq.~(\ref{eq:SWM}) in the basis of Hough harmonics yields prognostic equations for the complex Hough coefficients $\chi_{n,l}^k(t)$, which describe the evolution of amplitude and phase of each mode as
\begin{equation}
    \frac{d \chi_{n,l}^k(t)}{d t} + \left(i \nu_{n,l}^k+d_{n,l}^k\right) \chi_{n,l}^k(t) = f_{n,l}^k(t) + q_{n,l}^k(t).\label{eq:spectralproblem}
\end{equation}
Tendencies of $\chi_{n,l}^k$ are caused by the linear propagation, which depends on the eigenfrequency $\nu_{n,l}^k$ of the mode and the scale-selective viscosity $d_{n,l}^k$, which damps modes with large $k$ and $n$. The right-hand side of Eq.~(\ref{eq:spectralproblem}) contains the Hough transform of the interaction terms $\mathbf{N}(\lambda,\varphi,t)$, which is
\begin{equation}
    f_{n,l}^k(t) = \frac{1}{2\pi} \int_0^{2\pi} \int_{-\pi/2}^{\pi/2}\mathbf{N}(\lambda,\varphi,t)\cdot \left(\mathbf{H}_{n,l}^k(\lambda,\varphi)\right)^\ast \cos\varphi\, d\varphi\,d\lambda,\label{eq:expansion_nlin}
\end{equation} 
where the complex conjugate is denoted $^\ast$. Similarly, $q_{n,l}^k$ is the forcing in spectral space. Further details on the model numerics and time stepping are described in \citet{Vasylkevych.Zagar_2021}.

The non-dimensional energy of each mode is
\begin{equation}
    I_{n,l}^k = \frac{1}{2} (2-\delta_{0k})\left|\chi_{n,l}^k\right|^2, \label{eq:energy}
\end{equation}
where $\delta$ is the Kronecker delta.
The energy of all modes $I = \sum_{n,l,k} I_{n,l}^k$ is conserved in the linearized system without sources and sinks.
The dimensional energy is obtained by multiplying $I$ with $gD$, and it is equal to the sum in the domain of the kinetic energy and available potential energy of dimensional variables in physical space \citep{Kasahara.Puri_1981}.

The prognostic energy equation reads
\begin{equation}
    \frac{d I_{n,l}^k}{dt} = - 2d_{n,l}^kI_{n,l}^k + (2-\delta_{0k})\, \mathrm{Re}\left[ f_{n,l}^k \, \left({\chi_{n,l}^k}\right)^\ast + q_{n,l}^k \,\left({\chi_{n,l}^k}\right)^\ast\right],\label{eq:energy_tend}
\end{equation}
and it follows by differentiating Eq.~(\ref{eq:energy}) in time and using Eq.~(\ref{eq:spectralproblem}). The first term on the right-hand side of Eq.~(\ref{eq:energy_tend}) is the energy sink due to spectral viscosity. The other terms describe energy fluxes in modal space and due to the forcing as in \citet{Maho.etal_2024}.

\subsection{Wave-mean flow and wave-wave interactions}
\label{sect:wavemeanflow_wavewave}
In TIGAR, the backward transform of $\chi_{n,l}^k$ allows mode-selective filtering of the velocity components and depths associated with the modes and scales of interest. For filtering and diagnostics of our simulations, we split the modes into the following four groups: Rossby waves (denoted $R$), inertia-gravity waves and mixed Rossby-gravity waves, denoted $G$, Kelvin waves denoted $K$ and the zonal mean flow, which is denoted $0$. Note that our combination of IG and MRG waves into a single group ($G$) reflects their role in the present study, not their properties. The state vector thus consists of four parts, $\mathbf{X} = \sum \mathbf{X}_i$ for $i\in \{R, K, G, 0\}$. 

The interaction terms $\mathbf{N}$ consist of the quadratic terms in Eq.~(\ref{eq:SWM_physical}), which are advection, the metric terms, and the part of the divergence term which is not included in $\mathbf{L}$, as it contains fluid depth deviations from $D$. The total interactions are split into the interactions of different modes according to $\mathbf{N} = \sum \mathbf{N}_{i,j}$ for $(i,j)\in\{R, K, G, 0\}\times\{R, K, G, 0\}$, where
\begin{equation}
        \mathbf{N}_{i,j}   =\frac{-\gamma}{\cos\varphi} \begin{pmatrix}
         u_i \frac{\partial u_j}{\partial \lambda} + \upsilon_i \frac{\partial (u_j \cos\varphi)}{\partial \varphi}\\u_i \frac{\partial \upsilon_j}{\partial \lambda} + \upsilon_i \frac{\partial (\upsilon_j\cos\varphi)}{\partial \varphi} + (u_iu_j+\upsilon_i\upsilon_j) \sin\varphi\\u_i \frac{\partial h_j}{\partial \lambda} + \upsilon_i \cos\varphi \frac{\partial h_j}{\partial \varphi} + h_i \left( \frac{\partial u_j}{\partial \lambda} + \frac{\partial (\upsilon_j\cos\varphi)}{\partial \varphi}\right)
    \end{pmatrix}\,.
\label{eq:nlin_ij}
\end{equation}
The interactions of $i$ and $j$ with $i\neq j$ are $\mathbf{N}(i\leftrightarrow j) = \mathbf{N}_{i,j} + \mathbf{N}_{j,i}$.  The self-interactions of $i$ are $\mathbf{N}(i\leftrightarrow i) = \mathbf{N}_{i,i}$. Consequently, the terms involving waves and the mean flow are referred to as wave-mean flow interactions, and the products of wave terms are denoted wave-wave interactions. The energy flux due to the interactions is obtained using $f_{n,l}^k(i\leftrightarrow j)$ in Eq.~(\ref{eq:energy_tend}).
As triad interactions require that the sum of the zonal wavenumbers of the two interacting waves and the influenced mode must be zero \citep[e.g.][]{ripa_1982}, the interactions of waves ($k>0$) with the mean flow ($k=0$) do not affect the mean flow.

\subsection{Formulation of the background zonal mean state}
\label{sect:meanflow}
There are a few ways to configure a steady-state zonal mean flow in a barotropic fluid.
\citet{Kraucunas.Hartmann_2007} introduced a gradually-raising zonally symmetric topography centered at the equator, which leads to the development of symmetric zonal jets in the extratropics. A side product of the so produced steady state is the presence of meridional velocities \citep{barpanda_role_2023}. An alternative approach applied here and by \citet{Maho.etal_2024} is based on taking the steady state of the equation set (\ref{eq:SWM_physical}) with $\upsilon_0 = 0$. In this case the meridional momentum equation becomes
\begin{equation}
    u_0 + \frac{\gamma u_0^2}{\cos\varphi} + \frac{\gamma}{\sin\varphi} \frac{\partial h_0}{\partial\varphi} = 0\,. \label{eq:nonlinear_geostrophicbalance}
\end{equation}
Equation~(\ref{eq:nonlinear_geostrophicbalance}) defines the nonlinear balance, whereas the removal of the second term leaves the equation of the zonal geostrophic flow on the sphere (linear balance).
The Rossby modes with $k=0$ \citep{Kasahara_1978} fulfill the linear balance and are a part of the Hough harmonic computation in TIGAR. These modes have zero frequency and are used to compute a geostrophic zonal flow from a specified velocity profile $u_0(\varphi)$: The projection of $u_0+\gamma u_0^2/\cos\varphi$ on the zonal velocity components of Rossby modes with $k=0$ gives a set of coefficients. The linear combination of modes with these coefficients yields the non-linearly balanced depth field $h_0(\varphi)$.

\begin{figure}[t]
    \centering
    \includegraphics[width=.5\textwidth]{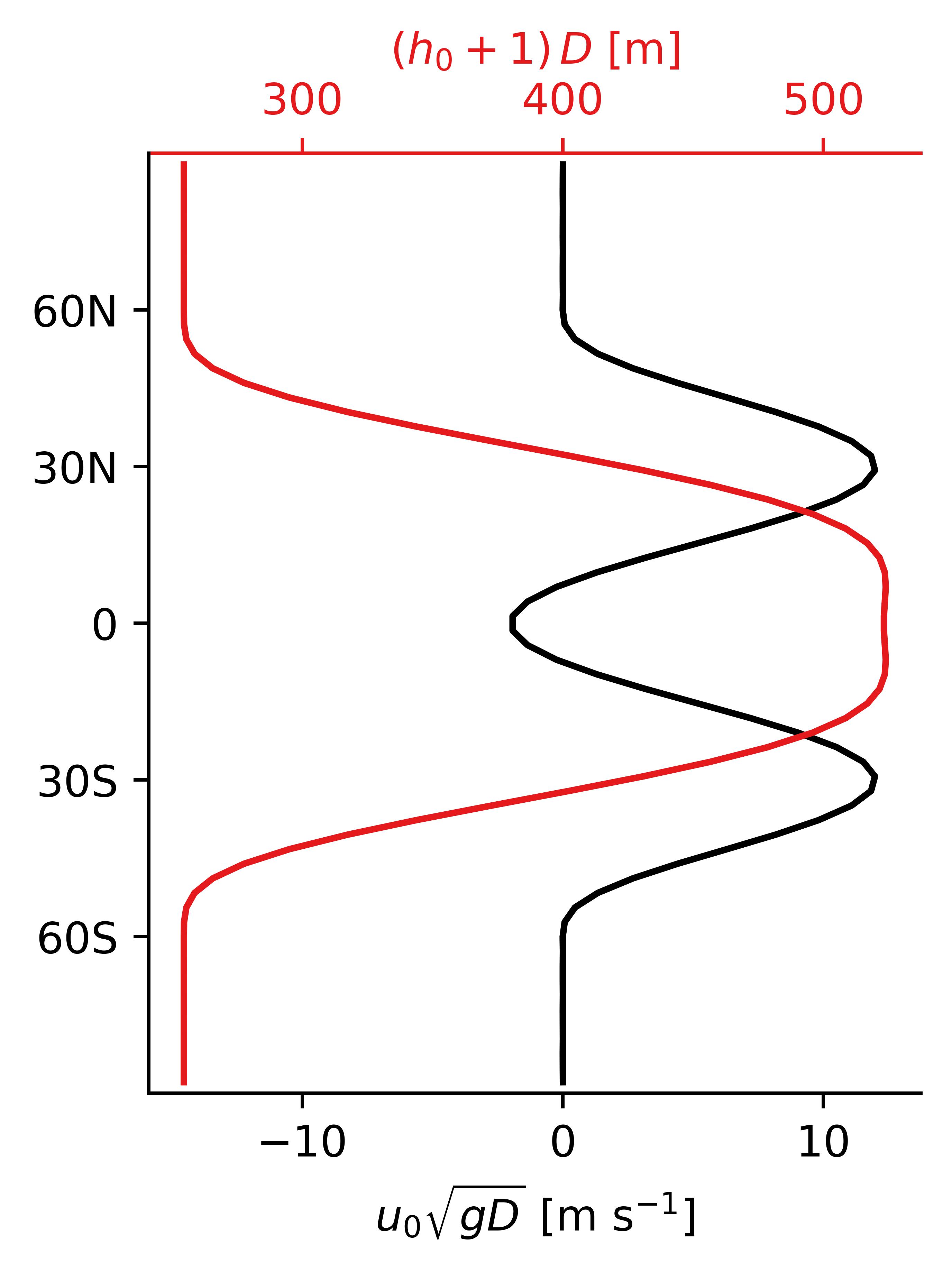}
    \caption{Zonal mean flow $u_0$ and fluid depth perturbation $h_0$ of the background zonal mean state.}
    \label{fig:zf4}
\end{figure}

Our balanced background state is shown in Fig.~\ref{fig:zf4}. The fluid has a mean depth of $D=400$~m, which corresponds to a vertical mode with baroclinic structure in the troposphere (a single zero crossing below the tropopause) and with the largest KW variance in the troposphere-stratosphere KW climatology in the ERA5 reanalysis \citep{Zagar.etal_2022}. The imposed zonal velocity profile resembles the upper-tropospheric climatology \citep[e.g.][]{Simmons_2022}, albeit with smaller amplitudes, which are a compromise between high velocities and moderate deviations from the mean depth. The profile is a symmetrical tenth degree polynomial, which is fitted to weak easterlies in the equatorial region and westerlies which assume their maximum at 30$^\circ$ off the equator. The zonal velocity profile is continuously differentiable and the velocity is zero poleward of 60$^\circ$. If the jet were narrower, as in \citet{Maho.etal_2024}, or closer to the equator, the balanced flow would have greater velocities for similar depth perturbations. However, the sensitivity to variations of the idealized background flow is beyond the scope of our study.

According to the stability criterion based on the potential vorticity \citep{ripa_1983}, the zonal flow in Fig.~\ref{fig:zf4} is stable since 
\begin{equation}
\frac{\partial PV}{\partial\varphi} \geq 0 \quad \mathrm{for}\,\,\mathrm{all} \,\,\,\varphi\quad \mathrm{and} \quad \mathrm{max}\left(\frac{u_0}{\cos\varphi}\right)\leq \mathrm{min}\left(\frac{u_0+\sqrt{h_0}}{\cos\varphi}\right)\,,
\end{equation}
where the potential vorticity of the mean flow is $PV = \frac{1}{h_0}\left(\sin\varphi - \frac{\gamma}{\cos\varphi} \frac{\partial(u_0\cos\varphi)}{\partial\varphi}\right)$.

The eigensolutions of Eq.~(\ref{eq:SWM_physical}), linearized around a non-resting balanced background state, have different eigenfrequencies and horizontal structures compared to the Hough harmonics, and we expect these differences to increase for stronger background flows. The numerical computation of these modified eigenmodes and their frequencies ${\nu'}_{n,l}^{k}$  is described in appendix A.  For the background flow in Fig.~\ref{fig:zf4}, the modified Rossby and Kelvin modes with $k=1$ are discussed in appendix B. The modified Kelvin mode and the modified Rossby modes with low $n$ are similar to the eigenmodes for the state of rest, in agreement with previous studies \citep{Boyd_1978a,Kasahara_1980_corr,Zhang.Webster_1989,Wang.Xie_1996,Mitchell_2013}. This supports our definition of the Rossby and Kelvin waves as the part of the circulation projecting on the Hough harmonics. The modified eigenmodes are not suitable to be used as basis functions of a model, as they are not an orthogonal basis (appendix A) and they vary in time for a changing background flow.

\begin{figure}[t]
    \centering
    \includegraphics[width=.5\textwidth]{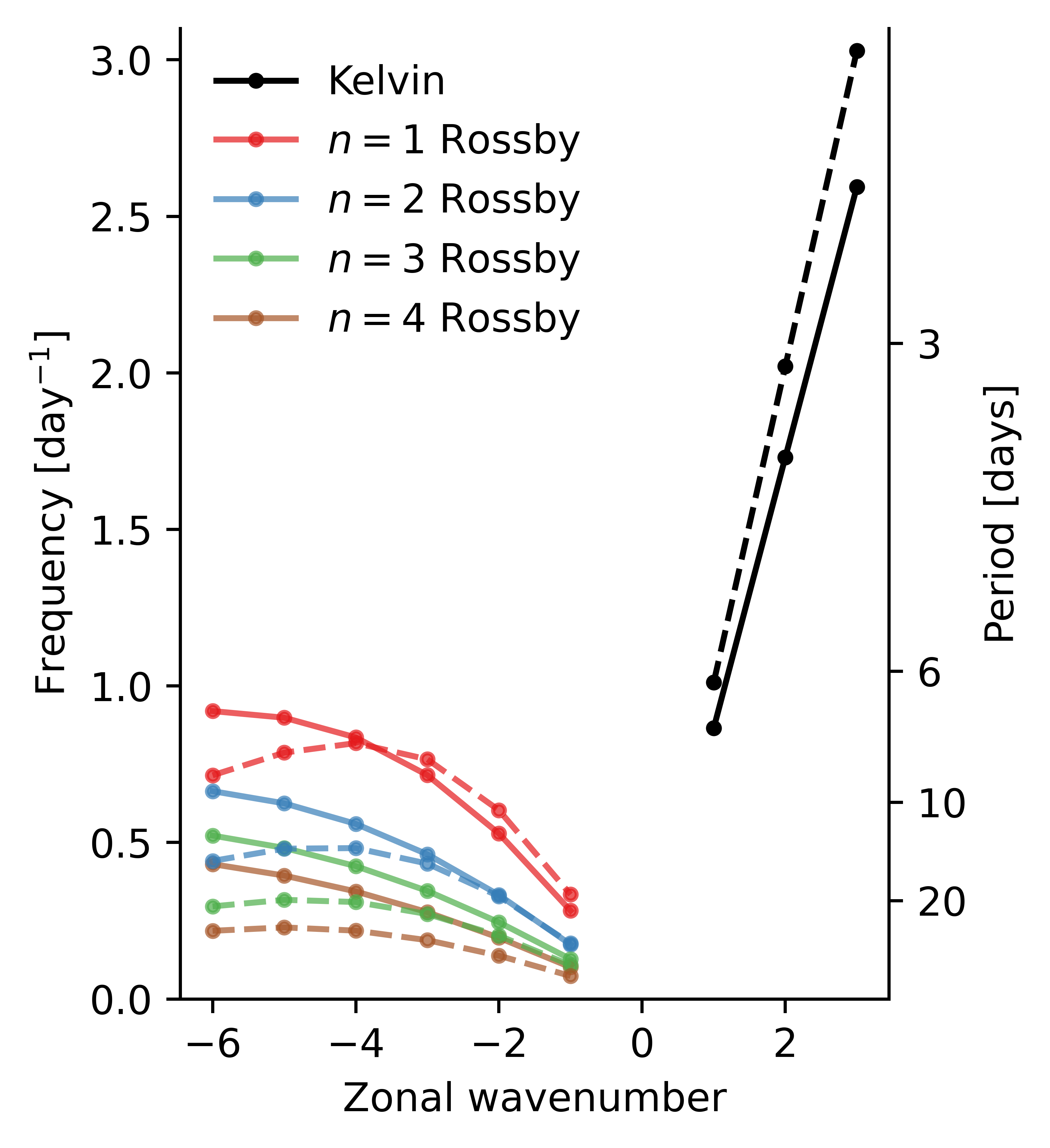}
    \caption{Dimensional frequencies of the Kelvin and Rossby waves on the sphere as a function of the zonal wavenumber. Solid lines: frequencies for the state of rest. Dashed lines: modified frequencies for the balanced background state shown in Fig.~\ref{fig:zf4}. The Rossby wave frequencies, which are negative, are shown as absolute values for negative $k$. The Rossby modes with $k=0$ have zero frequencies and there is no Kelvin $k=0$ mode.}
    \label{fig:zonalflow_all_kw}
\end{figure}

The modified frequencies include the Doppler shift by $u_0$, as well as effects of $h_0$. The comparison of the Kelvin and Rossby wave frequencies for the state of rest and the chosen background is presented in Fig.~\ref{fig:zonalflow_all_kw}. The weak equatorial easterlies decrease the KW frequency. However, the effect of the increased $h_0$ at the equator acts in the opposite sense so that the total effect is a somewhat greater KW frequency at each $k$. For example, the modified frequency of the KW with $k=1$ is ${\nu'}_K^1 = 1.01$~day$^{-1}(2\Omega)^{-1}$, which is higher than the unmodified frequency $\nu_K^1 = 0.86$~day$^{-1}(2\Omega)^{-1}$. The Rossby wave frequencies for small $k$ change relatively little.

The modified frequencies ${\nu'}_{n,l}^{k}$ are used to approximate the wave propagation in the presence of the background flow with analytical expressions. In the following, ${\nu'}_{n,l}^{k}$ denotes the modified frequencies if $n\leq 10$, and the frequencies of the Hough harmonics for $n>10$. This choice has no qualitative effect on the results because the modes with $n>10$ have small amplitudes in our simulations. In contrast, the frequencies of the Hough harmonics are used in  numerical simulations solving Eq.~(\ref{eq:spectralproblem}), where the effects of the background flow are included in the interaction terms.

\subsection{Subtropical Rossby wave forcing}
\label{sect:method_forcing}
We formulate a forcing that mimics Rossby waves advected by the subtropical jet in the upper troposphere. The speed of the westerly jet tends to be higher than the intrinsic westward Rossby wave phase speed, so that the Rossby waves are advected eastward \citep[and references therein]{Tulich.Kiladis_2021}. To simulate eastward-moving waves, the phase speed of the eastward-propagating forcing applied here is higher than $u_0$ of the relatively weak background flow in Fig.~\ref{fig:zf4}. The eastward advection of Rossby waves is thus conceptually represented by the external forcing.

While the eastward-propagating forcings employed by \citet{Zhang_1993} and \citet{Hoskins.Yang_2000} act on all modes of the system, the forcing in this study only affects Rossby waves. Such a forcing has a direct effect on the spectrum of the modified eigenmodes due to the non-resting background flow, including the modified Kelvin mode. However, the forcing only projects on the part of this modified mode that differs from the Kelvin wave. This difference accounts for the wave-mean flow interactions, which are included in the structure of the modified mode, because it is an eigensolution of the linearized equation (Eq.~\ref{eq:linearized}) containing these interactions. As the Rossby wave forcing does not affect the KW, Kelvin waves can only be excited by wave-mean flow and wave-wave interactions in our simulations. 

\begin{figure}[t]
    \centering
    \includegraphics[width=.5\textwidth]{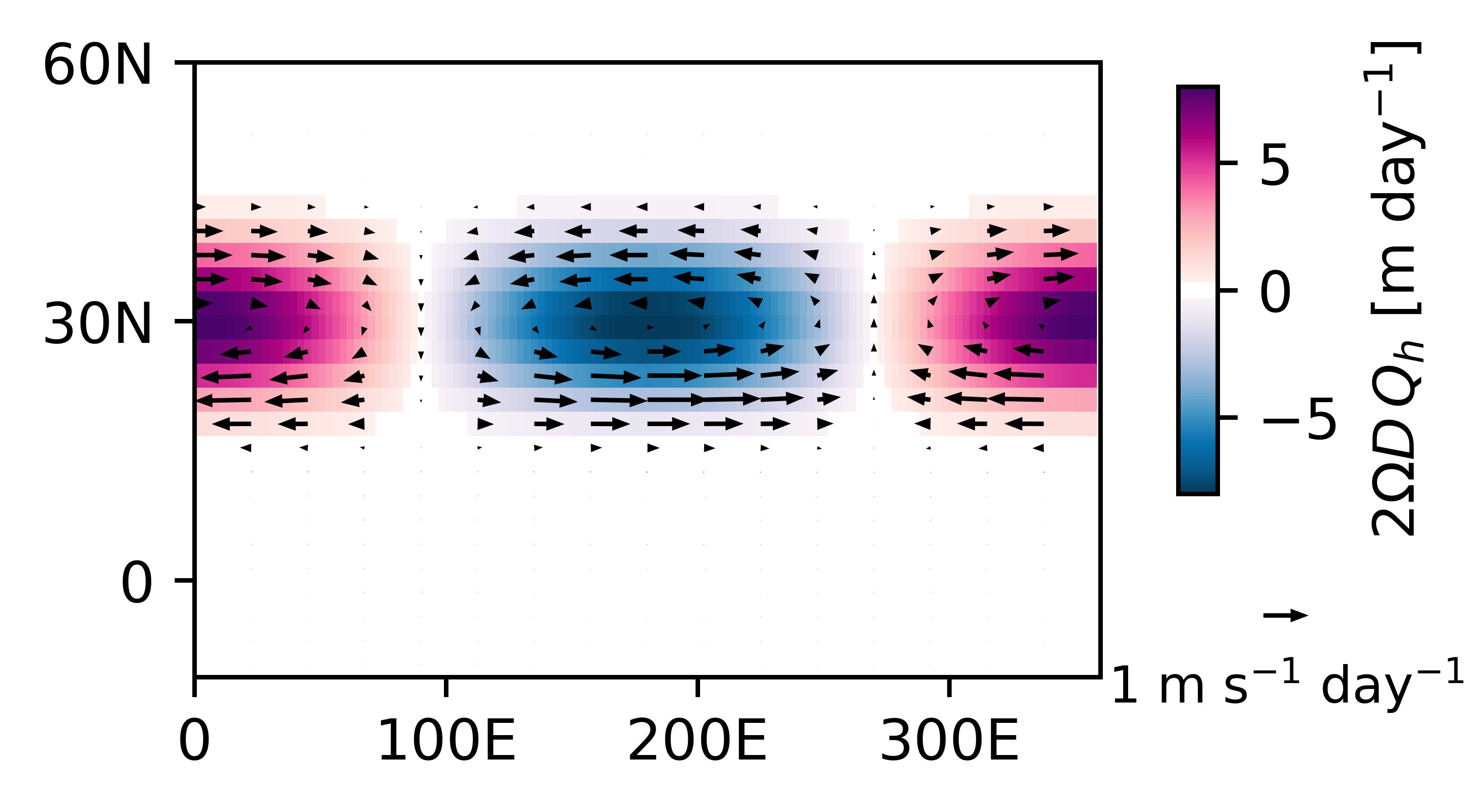}
    \caption{Horizontal structure of the Rossby wave forcing for $\varphi_0=30^\circ$ at the initial time. Shading denotes the forcing rate of the depth ($2\Omega D\, Q_h$), arrows stand for the momentum forcing rates ($2\Omega\sqrt{gD}\,(Q_u,Q_\upsilon)^\intercal$). }
    \label{fig:rossby_balanced}
\end{figure}

The forcing is restricted to $k=1$ and is defined as
\begin{equation}
    \mathbf{Q} = \sum_n \alpha_n\mathbf{\Theta}_{n,R}^1 e^{i(\lambda-\omega_0t)}\,.\label{eq:forcing}
\end{equation}
It propagates eastward if the forcing frequency $\omega_0$ is positive. The depth and velocity tendencies are located at the latitudes of subtropical westerlies (Fig.~\ref{fig:rossby_balanced}). The idealized meridional structure of the forcing is chosen to simulate eastward-moving Rossby waves as a general proof of concept.

In the following, we describe how the coefficients $\alpha_n$ are determined, so that the linear combination of global Rossby modes becomes a localized structure: An idealized depth perturbation profile 
\begin{equation}
    h(\varphi) = \cos^2(6(\varphi-\varphi_0)) \quad \mathrm{for}\quad \varphi \in (\varphi_0-15^\circ, \varphi_0+15^\circ)
\end{equation} is expressed as a linear combination of Rossby wave depth perturbations: $h(\varphi) = 2\sum_n\alpha_{n}' \, Z_{n,R}^1(\varphi)$ using a least-squares fit. From the coefficients $\alpha_n'$, the zonal Rossby wave velocities $u'(\varphi) = 2\sum_n\alpha_{n}' \, U_{n,R}^1(\varphi)$ are obtained. For the chosen $\varphi_0$ (section \ref{sect:method}\ref{sect:setup}), $u'(\varphi)$ contains zonal velocities at the equator, where the geostrophic balance does not apply, so that these velocities are not constrained by the prescribed depth perturbations. To localize the structure in the subtropics, we set $u'(\varphi)$ to zero for $\varphi \notin (\varphi_0-15^\circ, \varphi_0+15^\circ)$, which yields $u(\varphi)$. Finally, the coefficients $\alpha_n$ that fulfill $u(\varphi) = 2\sum_n\alpha_nU_{n,\mathrm{R}}(\varphi)$ are computed with another least-squares fit and scaled with 0.005 $\pi^{-1}$ so that the Rossby waves produced by the forcing are about one order of magnitude weaker than the mean state. For the chosen $\varphi_0$, the amplitudes of $\alpha_n$ are large for low meridional modes except for $n=1$ and close to zero for large $n$ (not shown).

\subsection{Simulation setup}
\label{sect:setup}
The numerical simulations use a triangular truncation of 42 modes ($T_{42}$) in spectral space and a Gaussian grid with $128\times64$ grid points in the longitudinal and latitudinal directions, respectively. The simulations are run for 40~days and the time step is 12~min. The exponential time differencing fourth-order Runge-Kutta method is chosen for the time stepping, see \citet{Vasylkevych.Zagar_2021} for further details. 

The initial condition is described in section~\ref{sect:method}\ref{sect:meanflow}, and the forcing according to section~\ref{sect:method}\ref{sect:method_forcing} is applied from the start of the simulations. In the reference simulation (REF),  $\varphi_0=30^\circ$, and the forcing frequency matches the frequency of the modified $k=1$ KW: $\omega_0 = {\nu'}_K^1 = 1.01$~day$^{-1}\,(2\Omega)^{-1}$. The sensitivity to the forcing frequency is investigated by the simulation LowF with a 20\% lower frequency. Additional sensitivity experiments are conducted in which the forcing frequency is the same as in REF and the central latitude of the forcing varies: Three simulations are performed with $\varphi_0=20^\circ$ (Q20), $\varphi_0=25^\circ$ (Q25), and $\varphi_0=35^\circ$ (Q35).

\section{Linear Rossby wave response to periodic forcing}
\label{sect:linrossby}
In the following, we describe the linear Rossby wave response to the forcing used in REF to facilitate the understanding of the nonlinear simulations described in section~\ref{sect:results}.
The linear response to the forcing (Eq.~\ref{eq:forcing}) is the solution of
\begin{equation}
    \frac{d\chi_n}{dt} + (i\nu_n'+d_n)\chi_n = \alpha_n\,e^{-i\omega_0t}\,,
    \label{eq:rw_lin_prognostic}
\end{equation}
where $\chi_n$ is the Hough expansion coefficient for the $n$th meridional Rossby mode with $k=1$. To approximate the effects of the mean flow on the wave propagation, Eq.~(\ref{eq:rw_lin_prognostic}) involves the modified frequency as discussed in section~\ref{sect:method}\ref{sect:meanflow}. As the spectral viscosity only affects the waves with large wavenumbers, we can assume that $d_n=0$ in the following analytical calculations. The solution of Eq.~(\ref{eq:rw_lin_prognostic}) is well known \citep[e.g.][]{Kasahara_1984,Zhang_1993}:
\begin{subequations}
\begin{eqnarray}
    \chi_n(t) &=& \alpha_ne^{-i\nu_n't}\int_0^t e^{i(\nu_n'-\omega_0)t'}dt' \label{eq:kasahara84_4.7}\\
    &=& \frac{i\alpha_n}{\nu'_n-\omega_0}\left(e^{-i\nu'_nt}-e^{-i\omega_0t}\right)\,,
    \label{eq:hcoeff_forced_rw}
\end{eqnarray}
\end{subequations}
and it consists of two summands: the free solution of the homogeneous equation and the forced solution of the inhomogeneous part of Eq.~(\ref{eq:rw_lin_prognostic}).

\begin{figure}[t]
    \centering
    \includegraphics[width=
\textwidth]{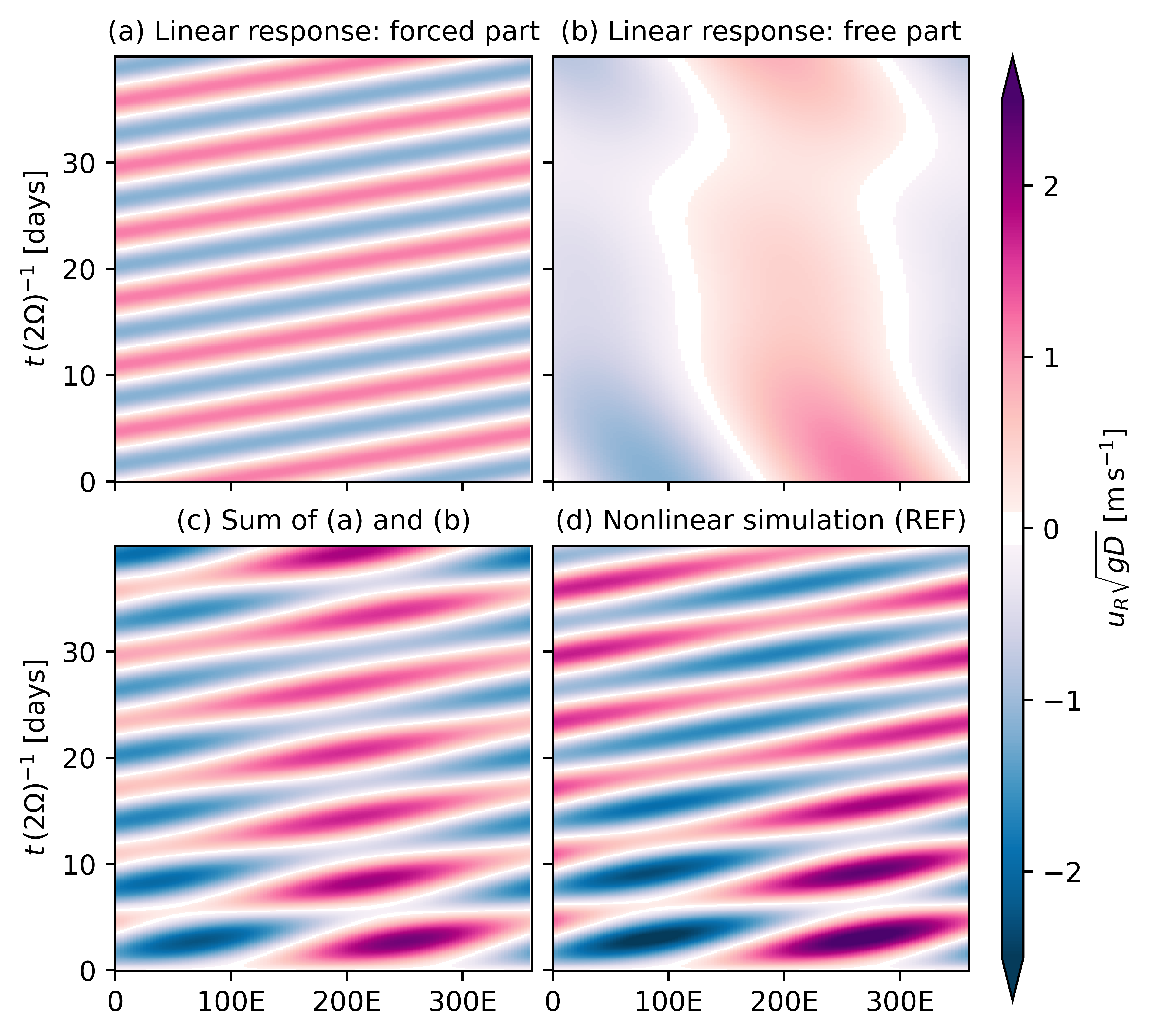}
    \caption{Rossby wave zonal velocity anomalies at $20^\circ$N, excited in response to the forcing of Fig.~\ref{fig:rossby_balanced}. (a) forced and (b) free part of the linear response (Eq.~\ref{eq:hcoeff_forced_rw}), and (c) their sum, (d) nonlinear reference simulation.}
    \label{fig:ref_energy_ros_analytical}
\end{figure}

Transforming each summand in Eq.~(\ref{eq:hcoeff_forced_rw}) to physical space yields the velocities and depth perturbations of the free and forced Rossby waves. The evolution of the zonal velocity for the two components is shown in Fig.~\ref{fig:ref_energy_ros_analytical} at $20^\circ$N, which is close to the latitude of the maximum velocity forcing (Fig.~\ref{fig:rossby_balanced}).

The forced Rossby waves (inhomogeneous part of the solution) propagate eastward (Fig.~\ref{fig:ref_energy_ros_analytical}a), with their phase velocity determined by the choice of $\omega_0$.
The free Rossby wave solution (Fig.~\ref{fig:ref_energy_ros_analytical}b) is a superposition of waves with several $n$ with different frequencies $\nu_n'$. These frequencies are negative for the meridional modes with the largest amplitudes, for which $n$ is small. As the phases of the Rossby waves with different $n$ vary during the simulation, the magnitude of the zonal velocities of the free Rossby waves decrease with time. After about 30 days, the zonal velocities increase again as the phases of different modes align again (Fig.~\ref{fig:ref_energy_ros_analytical}b). 

In the superposition of the free and forced Rossby waves (Fig.~\ref{fig:ref_energy_ros_analytical}c), the eastward propagation is clearly recognizable, and superposed on it are variations with a period of about 6~days.
The variations also appear in the energy of the $k=1$ Rossby waves calculated using Eq.~(\ref{eq:energy}) and Eq.~(\ref{eq:hcoeff_forced_rw}):
\begin{equation}
    I_n(t)= \frac{4\alpha_n^2}{(\nu'_n-\omega_0)^2}\sin^2\left(\frac{\nu'_n-\omega_0}{2}t\right)\,.
\label{eq:energy_rossby_basic_analytical}
\end{equation}
The maxima of the $\sin^2$ function in Eq.~(\ref{eq:energy_rossby_basic_analytical}) are about 5-6 days apart for small $n$ (not shown).

In summary, the external forcing generates Rossby waves in the subtropics, which move eastward and vary periodically. Differences between the linear response and the nonlinear simulations will be described in the next section.

\section{Kelvin wave excitation}
\label{sect:results}
Now we describe the excitation of KWs in the nonlinear simulations. The KW energy tendencies are then attributed to wave-wave and wave-mean flow interactions. Subsequently, we explain the major part of the KW excitation as a resonance mechanism with interactions between the mean flow and the forced Rossby waves.

\subsection{Nonlinear simulations}
\label{sect:simulation}
\begin{figure}[t]
    \centering
    \includegraphics[width=
.5\textwidth]{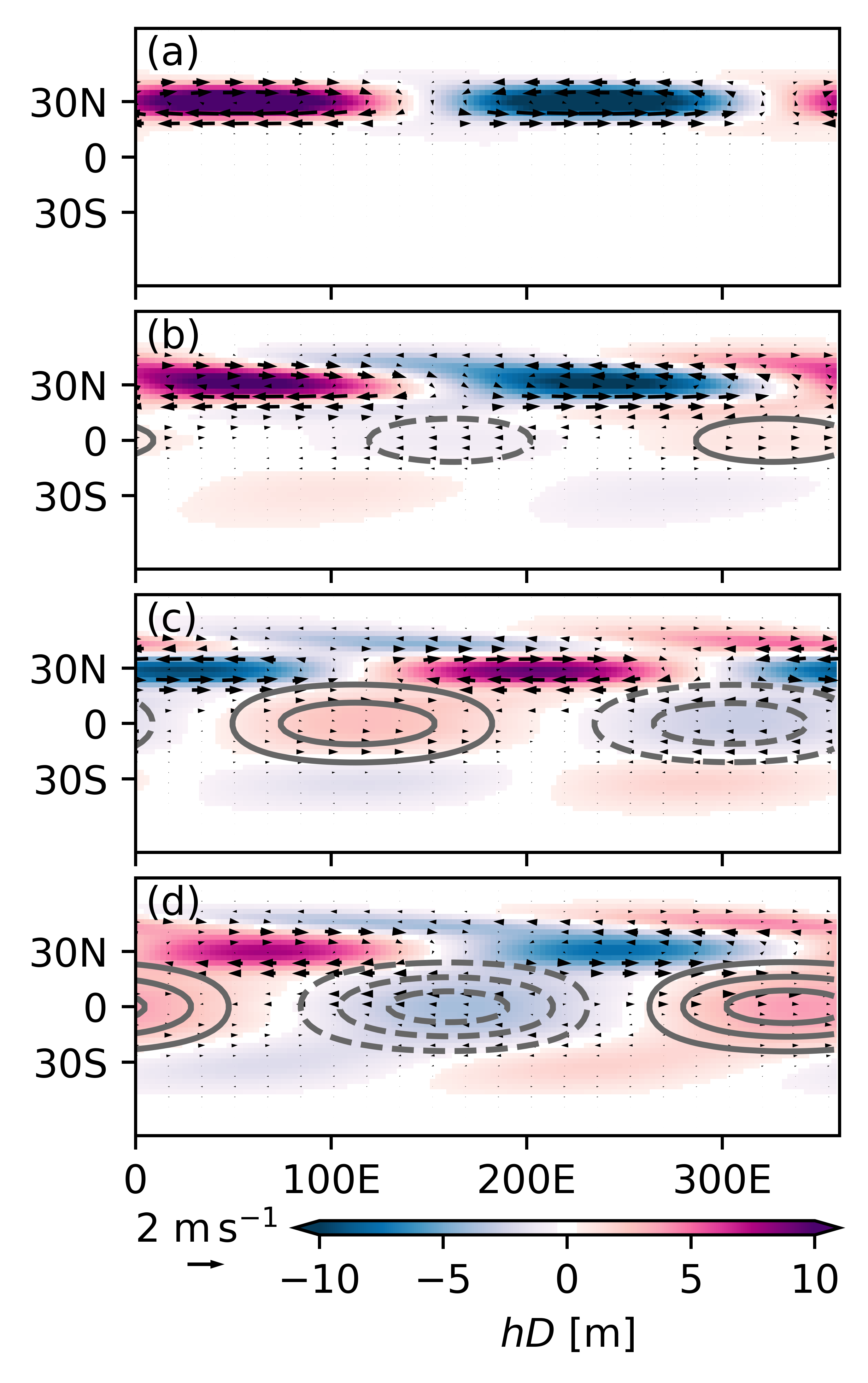}
    \caption{Fluid depth perturbations $h\,D$ (shading) and velocity perturbations $(u,v)^\intercal\,\sqrt{gD}$ (arrows) in the REF simulation after (a) 2 days, (b) 15 days, (c) 30 days and (d) 40 days. The KW depth perturbation $h_K\,D$ is shown by gray contours with line spacing of $\pm\,1$~m. Solid and dashed lines denote positive and negative depth perturbations, respectively.}
    \label{fig:ref_horiz}
\end{figure}

In REF, the Rossby waves generated by the external forcing are initially in phase with the forcing. As their amplitudes grow, their phases gradually shift eastward (Fig.~\ref{fig:ref_horiz}a). The initial growth of the Rossby waves is part of their periodic variation, which is similar to the linear response (Fig.~\ref{fig:ref_energy_ros_analytical}c and d). Rossby waves also appear as velocity and depth perturbations in the southern-hemispheric subtropics and north of the forcing, because several global modes, which have different phase velocities, are affected by the forcing (section \ref{sect:method}\ref{sect:method_forcing}). After 15 days, the KW with $k=1$ is evident in the equatorial region (Fig.~\ref{fig:ref_horiz}b), and it continues to grow throughout the simulation (Fig.~\ref{fig:ref_horiz}c,d). The depth perturbations at the equator are almost entirely composed of the $k=1$ KW. The Kelvin waves have a fixed phase lag relative to the Rossby waves around 30$^\circ$ North. We refer to the KW growth as "excitation" because no KWs are present at the beginning of the simulation.

\begin{figure}[t]
    \centering
    \includegraphics[width=.5
\textwidth]{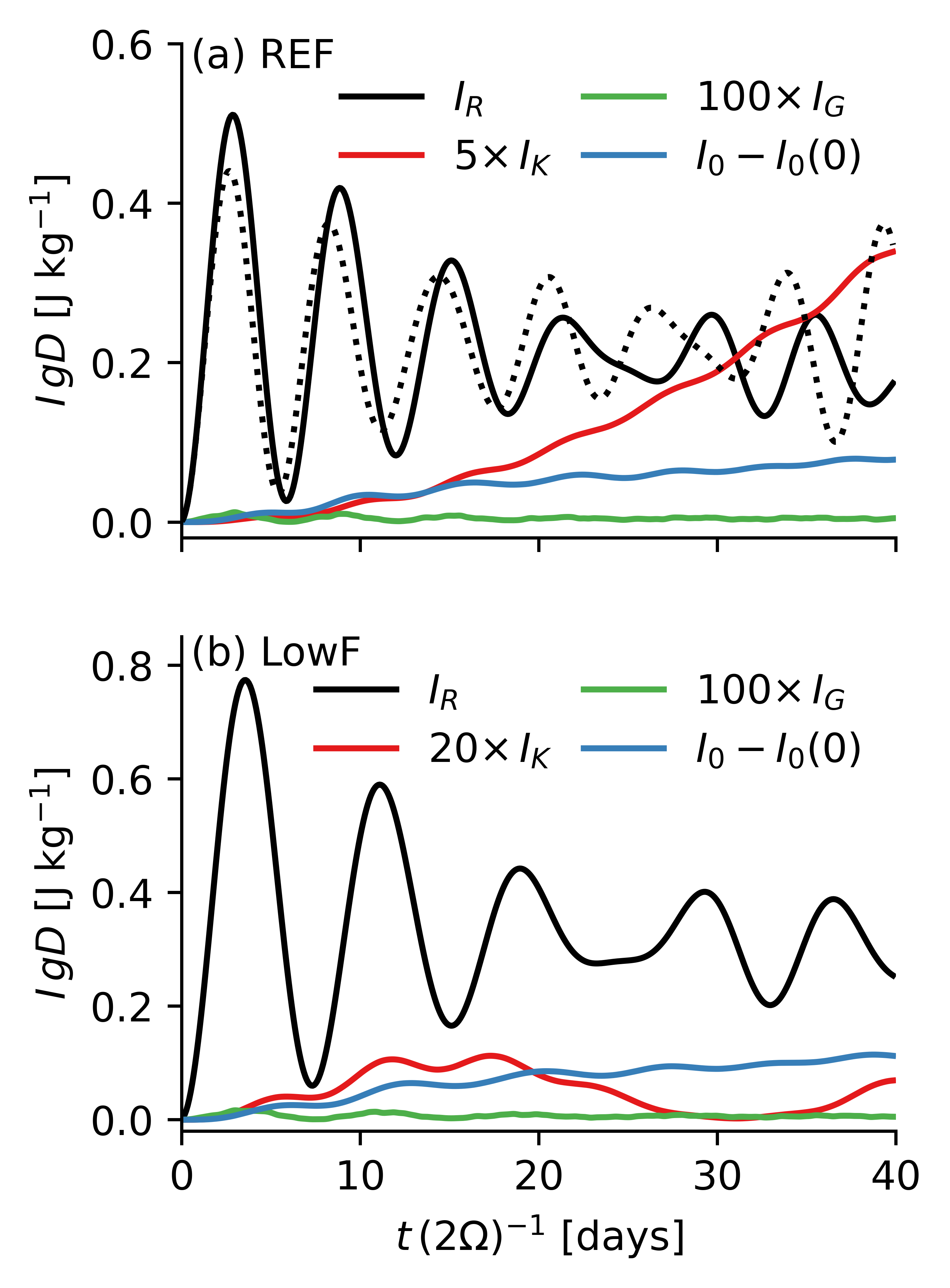}
    \caption{Evolution of energy in nonlinear simulations. (a) REF and (b) LowF simulations. The energy is split among Rossby waves $I_R$, Kelvin waves $I_K$, inertia-gravity waves $I_G$ and the zonal mean flow $I_0$. For waves, the energy is summed over all $k$. The dotted line in (a) is the solution for the linear case (Eq.~\ref{eq:energy_rossby_basic_analytical}). Note that energies are multiplied by a factor of 5 and 100 in (a) and 20 and 100 in (b) for $I_K$ and $I_G$, respectively.}
    \label{fig:energy}
\end{figure}

The KW energy evolution is shown in Fig.~\ref{fig:energy}, along with the energy variations in other components of the flow. In REF (Fig.~\ref{fig:energy}a), the KW energy grows with time and after one month it reaches about 20\% of the Rossby wave energy. The energy of the mean flow slowly increases with time, while the Rossby wave energy oscillates due to the external forcing.
This variation is similar to the linear response, but the period and amplitude of the energy variation differ (dotted versus full black lines in Fig.~\ref{fig:energy}a). One reason for this mismatch is the meridional variation of the mean flow: the strong westerlies slow down the westward Rossby wave propagation in the subtropics, so that the spatial structure of the Rossby waves is modified. Such modifications of waves by wave-mean flow interactions, which also appear in the modified structure of the modes in the presence of the background flow, are not considered in the linear response to the forcing. Due to the resting background state, such interactions are precluded there by design, and only the frequencies of the modes are changed. Additionally, the linear response does not include wave-wave interactions and temporal variations of the mean flow. In total, these processes slow the growth of the phase difference between the Rossby waves and the forcing. As a result, the period of the Rossby wave energy oscillation in REF is longer than in the linear response, and the mismatch increases over time (Fig.~\ref{fig:energy}a). The energy of the IG and MRG waves is close to zero.

In the simulation with lower forcing frequency (LowF), the KW energy is considerably lower than in REF, and it oscillates with a period of about 30 days (Fig.~\ref{fig:energy}b). For longer simulation times, the periodic behaviour in LowF continues, while the KW energy growth in REF stagnates much later (after about 120 days, not shown). This difference implies that the forcing frequency determines whether the KW energy grows continuously or oscillates with time. The energies of $G$ and the mean flow in LowF are similar to REF. The Rossby wave energy in LowF oscillates with a longer period and greater amplitude than in REF (Fig.~\ref{fig:energy}b), which is in line with Eq.~(\ref{eq:energy_rossby_basic_analytical}) where the absolute value of $\nu_n'-\omega_0$ is smaller when $\omega_0$ is decreased by 20\%.

The relatively small KW energy level in REF differs from the simulations by \citet{Hoskins.Yang_2000}, where the Kelvin wave amplitudes were similar to the Rossby wave amplitudes. This difference might be due to the direct effect of their heating or vorticity forcing on the Kelvin waves, among further differences in the model and setup.

In conclusion, the energy growth of the KWs depends on the frequency of the Rossby wave forcing. In the following subsections, we explain details of the KW excitation process. To better understand the initial excitation, we  combine numerical simulations with the analytical solution, for which the Rossby waves are approximated with the linear response.

\subsection{Wave-mean flow and wave-wave interactions}
\label{sect:nlin_sim}
The KW energy tendency is caused by the quadratic terms, because the linear response to the external forcing only includes Rossby waves, and the spectral viscosity does not influence waves with small $k$. Using the notation introduced in Eq.~(\ref{eq:nlin_ij}) for interaction terms and neglecting the dissipation due to spectral viscosity, the energy tendency equation (Eq.~\ref{eq:energy_tend}) for the KWs becomes 
\begin{subequations}
    \begin{eqnarray}
        \frac{dI_K}{dt} &=& 2(\mathrm{Re}[f_K(R\leftrightarrow 0)\chi_K^\ast] + \mathrm{Re}[f_K(G\leftrightarrow 0)\chi_K^\ast] + \mathrm{Re}[f_K(R\leftrightarrow R)\chi_K^\ast] \label{eq:kwtendsplita}\\
        &+& \mathrm{Re}[f_K(R\leftrightarrow G)\chi_K^\ast] + \mathrm{Re}[f_K(K\leftrightarrow K)\chi_K^\ast] + \mathrm{Re}[f_K(K\leftrightarrow G)\chi_K^\ast]\label{eq:kwtendsplitb}\\
        &+& \mathrm{Re}[f_K(K\leftrightarrow 0)\chi_K^\ast] + \mathrm{Re}[f_K(G\leftrightarrow G)\chi_K^\ast] + \mathrm{Re}[f_K(R\leftrightarrow K)\chi_K^\ast]) \label{eq:kwtendsplitc}\,.
    \end{eqnarray}
    \label{eq:kwtendsplit}
\end{subequations}
Equation~(\ref{eq:kwtendsplit}) involves all $k>0$ as there is no KW with $k=0$, and $f_K(0\leftrightarrow 0)$ is thus omitted.

\begin{figure}[t]
    \centering
    \includegraphics[width=.5
\textwidth]{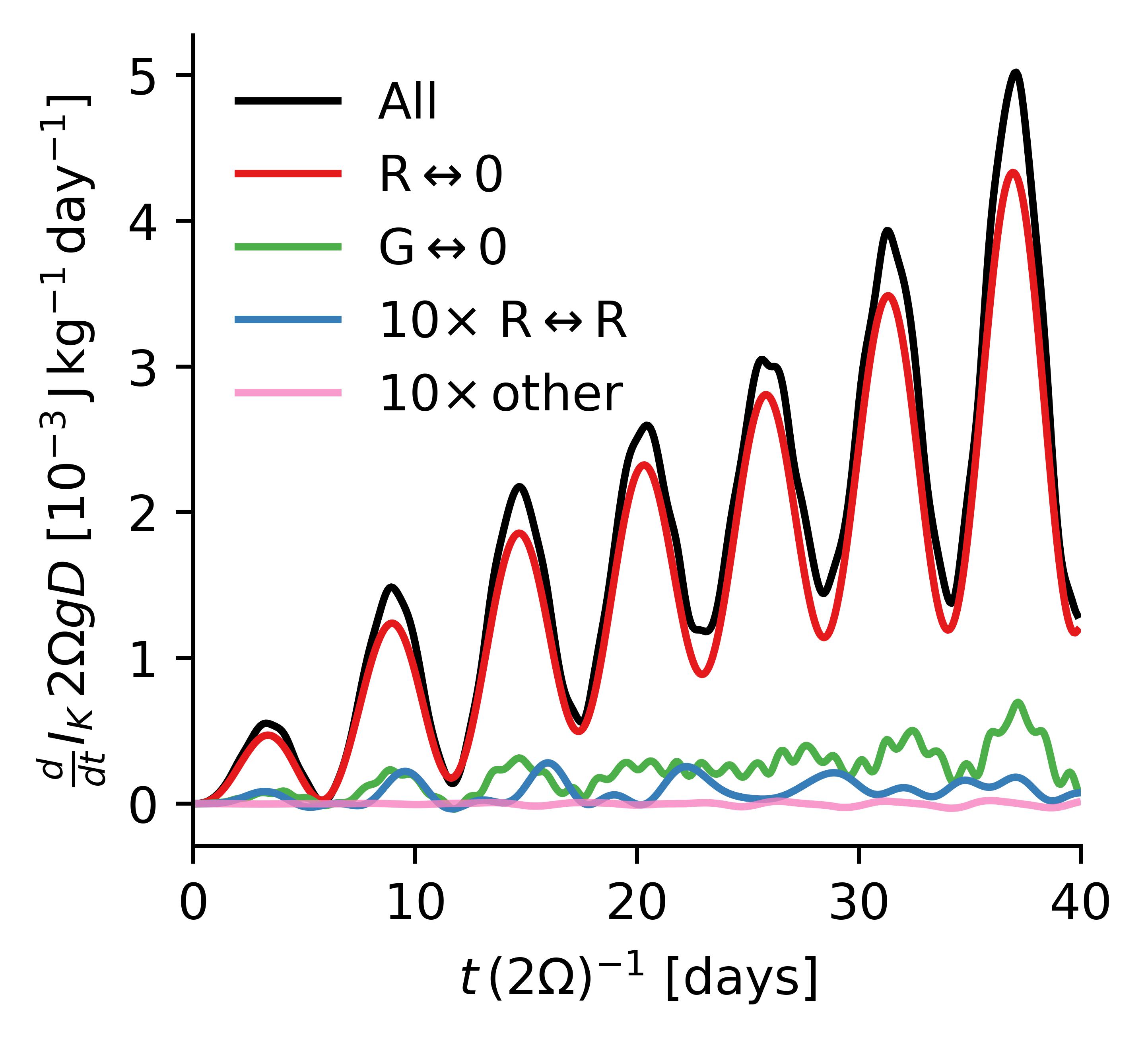}
    \caption{KW energy tendencies (All) split into contributions due to Rossby wave-mean flow interactions ($R\leftrightarrow 0$), gravity wave-mean flow interactions ($G\leftrightarrow 0$), Rossby wave-wave interactions ($R\leftrightarrow R$) and all other wave-wave and wave-mean flow interactions (Eq.~\ref{eq:kwtendsplit}). The sum over all zonal wavenumbers is shown.}
    \label{fig:ref_energytend}
\end{figure}

The various contributing terms are presented for the REF simulation in Fig.~\ref{fig:ref_energytend}. It shows that the total KW energy tendency due to all interactions, $\frac{dI_K}{dt}, $ is positive, which indicates that the interactions are roughly in phase with the Kelvin wave. The KW energy tendency oscillates with time due to the periodicity of the Rossby wave amplitude discussed in section \ref{sect:linrossby}. About 90\% of the KW energy tendency is caused by the Rossby wave-mean flow interactions ($R\leftrightarrow 0$). The second-largest tendency contribution is due to the gravity wave-mean flow interactions ($G\leftrightarrow 0$), which contribute about 10\%. The Rossby wave-wave interactions ($R\leftrightarrow R$) are smaller than $R\leftrightarrow 0$ and cause less than 1\% of the KW energy growth. The amplitude of the interactions depends on the magnitude of the depth perturbations and velocities of the interacting modes, which are smaller for the Rossby waves than for the mean flow. The other interactions (Eq.~\ref{eq:kwtendsplitb} and \ref{eq:kwtendsplitc}) cause negligibly small KW energy tendencies (Fig.~\ref{fig:ref_energytend}), because they contain waves with small amplitudes.

The dominant zonal wavenumber of the Rossby wave-mean flow interactions is $k=1$, which is the sum of the zonal wavenumbers of the Rossby waves and the mean flow. The zonal wavenumber of the KW excited by $R\leftrightarrow 0$ is thus $k=1$ (Fig.~\ref{fig:ref_horiz}). Accordingly, the Rossby wave-wave interactions generate KWs with $k=2$, whose energy is two orders of magnitude smaller than the $k=1$ KW energy. KWs with higher $k$ have even lower amplitudes because they are only excited by higher-order interactions, which are small.

The described results differ from \citet{Cheng.etal_2022}, who explained the KW excitation with eddy momentum fluxes, i.e. wave-wave interactions, while the wave-mean flow interactions damped the KWs. However, our experiments do not necessarily contradict \citet{Cheng.etal_2022} who identified synoptic-scale Rossby waves in the subtropics as relevant for the KW growth and accordingly found $k=5$ KWs excited by subtropical dynamics.

The dominant contribution of the Rossby wave-mean flow interactions to the KW excitation with peak signal at $k=1$ is found in all sensitivity simulations with varying forcing latitudes (not shown). In LowF, the KW energy decreases after its initial growth because the phase of the Rossby wave-mean flow interactions shifts relative to the KWs. After about 30 days, the phases are aligned and the KW energy grows again. In REF, Q20, Q25 and Q35, however, the Rossby wave-mean flow interactions are always in phase with the KWs so that their energy grows continuously. In a longer run, the mismatch between the forcing frequency and the frequency of the modified Kelvin mode increases, because the modified eigenmodes gradually change along with the mean flow. This leads to a stagnation of the KW growth.
In the following, we focus on the dominant process, the Rossby wave-mean flow interactions.

\subsection{Resonant Kelvin wave excitation}
\label{sect:resonance}
The Rossby wave-mean flow interactions contain interactions of both the forced and the free Rossby waves with the mean flow. Since the KW excitation is sensitive to the forcing frequency (section~\ref{sect:results}\ref{sect:simulation}), the interactions of the forced Rossby waves with the mean flow presumably cause a greater proportion of the energy growth than the free Rossby wave-mean flow interactions. Now we explain how the resonant KW excitation by these interactions takes place.

The forced Rossby wave-mean flow interactions excite KWs with $k=1$, whose expansion coefficient evolves according to
\begin{equation}
    \frac{d\chi_K}{dt} + i \nu_K' \chi_K = \hat{f} e^{-i\omega_0t}.
    \label{eq:kw_prognostic_forced}
\end{equation}
The forced Rossby wave-mean flow interactions propagate eastward as determined by $\omega_0$. The KW component of their Hough transform, evaluated at the initial time, is denoted $\hat{f}$. Equation~(\ref{eq:kw_prognostic_forced}) neglects the interactions of the free Rossby waves with the mean flow, the interactions of the IG, MRG and Kelvin waves with the mean flow, wave-wave interactions, as well as magnitude changes of the forced Rossby wave-mean flow interactions due to the temporally varying mean flow.

\begin{figure}[t]
    \centering
    \includegraphics[width=.5
\textwidth]{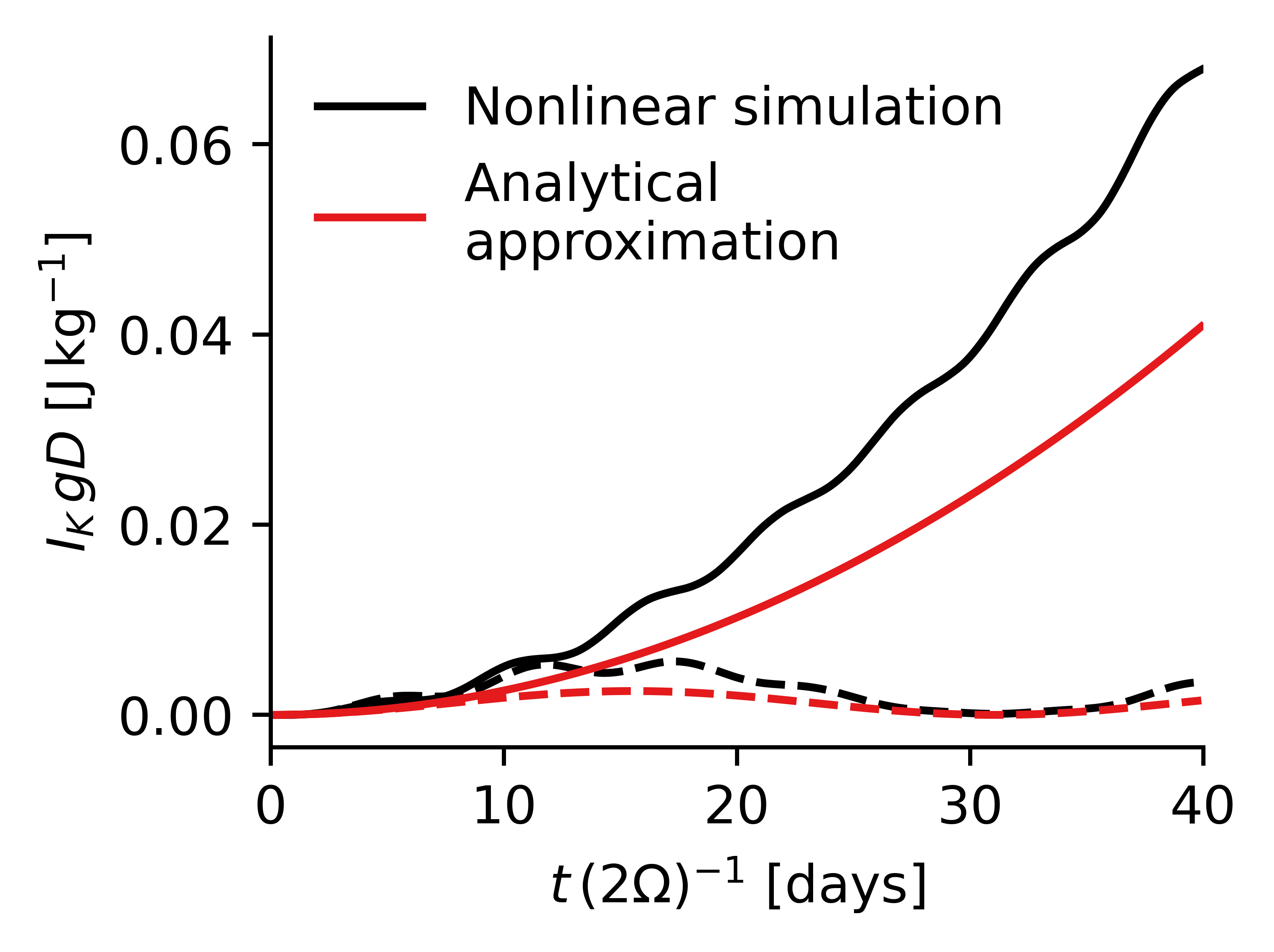}
    \caption{Evolution of the $k=1$ KW energy in the nonlinear simulations REF (solid black) and LowF (dashed black), and according to the analytical approximations (red lines) where the energy is changed only by the forced Rossby wave-mean flow interactions (Eqs.~\ref{eq:energy_kw_resonance}-\ref{eq:energy_kw_nonresonance}).}
    \label{fig:kw_lin_nlin}
\end{figure}

The solution of Eq.~(\ref{eq:kw_prognostic_forced}) is 
\begin{equation}
    \chi_K = -i\hat{f}e^{-i\nu_K't}\,\frac{e^{i(\nu_K'-\omega_0)t}-1}{\nu_K'-\omega_0}.
    \label{eq:hcoeff_forced_kw}
\end{equation}
In the limit of $\omega_0$ approaching $\nu_K'$, the L'H\^opital rule applies and the KW amplitude $|\chi_K|$ increases linearly. In this resonance case, the KW energy grows perpetually (Fig.~\ref{fig:kw_lin_nlin}):
\begin{equation}
    I_K(t) = \hat{f}^2t^2 .\label{eq:energy_kw_resonance}
\end{equation}
If the forcing frequency deviates from the KW eigenfrequency, the KW energy oscillates periodically according to
\begin{equation}
    I_K(t) = \frac{4\hat{f}^2}{(\nu_K'-\omega_0)^2}\sin^2\left(\frac{\nu_K'-\omega_0}{2}t\right). \label{eq:energy_kw_nonresonance}
\end{equation}
The period of this oscillation increases as $\omega_0$ approaches $\nu_K'$. In LowF, the period is 31~days.

The evolution of the KW energy in REF and LowF and the analytical approximation is shown in Fig.~\ref{fig:kw_lin_nlin}.
It confirms that the major part of the KW energy growth in REF is due to resonance with the forced Rossby wave-mean flow interactions (solid lines in Fig.~\ref{fig:kw_lin_nlin}). The same applies to LowF, but the KW energy level is much smaller.
In addition, the KW energy in the nonlinear simulations varies with a period of about 6~days (Fig.~\ref{fig:kw_lin_nlin}). This is the period of the Rossby wave energy oscillation in Fig.~\ref{fig:energy}, which includes the free waves. Moreover, the simulated KW energy is greater than the analytical approximation because interactions of IG waves with the mean flow and wave-wave interactions (Fig.~\ref{fig:ref_energytend}) also contribute to the KW growth. 
Nevertheless, the analytical approximation explains the qualitative behaviour seen in Fig.~\ref{fig:kw_lin_nlin} well, especially during initial stages of the simulations.

In addition to Kelvin waves, other modes can also be resonantly excited if $\omega_0$ is close to their frequencies, and if other factors such as the meridional shear of the zonal flow result in a nonzero Hough transform of the wave-mean flow or wave-wave interactions for these modes. For instance, the meridional shear is essential for the energy exchange between extratropical barotropic and tropical baroclinic Rossby waves studied by \citet{majda_biello_2003}. Furthermore, interactions of a barotropic flow with meridional shear and baroclinic Kelvin waves can excite other equatorial waves \citep{Ferguson.etal_2009}.

The modification of the eigenmodes by the non-resting background flow (see appendix) offers a complementary interpretation of the KW excitation by Rossby wave-mean flow interactions. The forcing directly excites the modified Kelvin mode, and it is in phase with this mode because of the chosen forcing frequency (sections \ref{sect:method}\ref{sect:method_forcing} and \ref{sect:method}\ref{sect:setup}). Therefore, the amplitude of the modified Kelvin mode increases, which is diagnosed as Kelvin wave energy growth.

\subsection{Tendencies due to forced Rossby wave-mean flow interactions in physical space}
\label{sect:rw_bg_int}
Now we analyze tendencies due to forced Rossby wave-mean flow interactions in physical space to separate the effects on the velocity and depth variables. We first discuss the tendencies for REF, and then explore their sensitivity to the location of the forcing for Q20, Q25 and Q35. The latitudinal variation of the mean flow influences the meridional structure of the forced Rossby wave-mean flow interactions, so that they impact various modes, including the KW. The analysis is shown for a single time step and longitude. The forced Rossby waves have a constant amplitude and they maintain a fixed phase shift relative to the KWs if their frequencies match. The impact on the KWs thus changes little over time, as the mean flow is approximately constant.

\begin{figure}[t]
    \centering
    \includegraphics[width=
\textwidth]{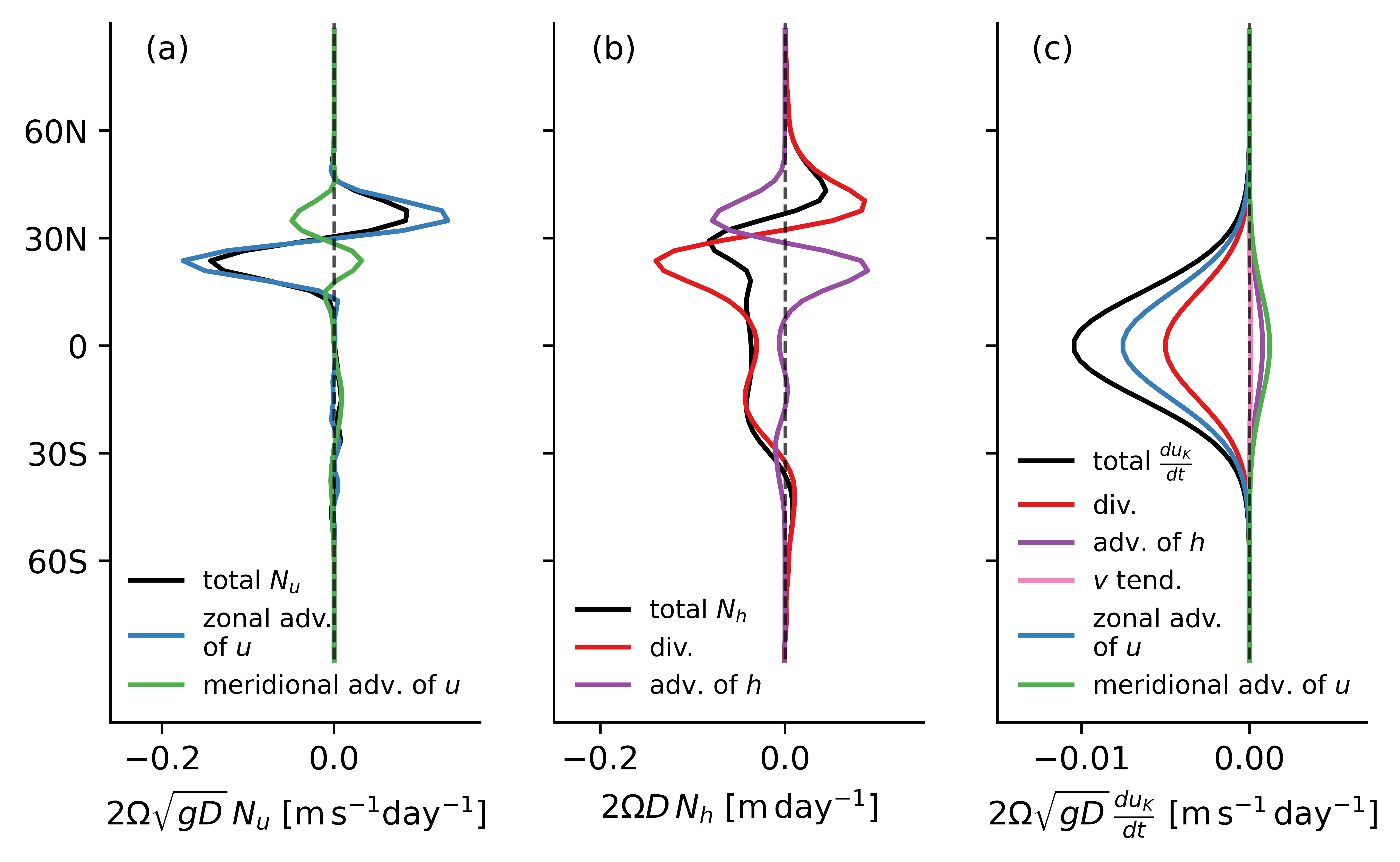}
    \caption{(a)-(b) Interactions of forced Rossby waves and the mean flow in REF at $t=0$ and $0^\circ$E. Tendency of the (a) zonal velocity ($N_u$, Eq.~\ref{eq:tend_u}), and (b) fluid depth ($N_h$, Eq.~\ref{eq:tend_z}). (c) $k=1$ KW zonal velocity tendency $\frac{du_K}{dt}$ due to each term in (a) and (b) and the meridional velocity tendencies.}
    \label{fig:tend_uZ}
\end{figure}

In the following, $R$ denotes the forced Rossby waves, which have $k=1$.
The contribution of the forced Rossby wave-mean flow interactions to the zonal velocity tendency
\begin{equation}
    N_u = \underbrace{\frac{-\gamma}{\cos\varphi}u_0\frac{\partial u_R}{\partial \lambda}}_\mathrm{zonal\,\, advection} \underbrace{- \gamma \upsilon_R\frac{\partial( u_0\cos\varphi)}{\partial \varphi}}_\mathrm{meridional\,\, advection} \label{eq:tend_u}
\end{equation}
consists of two advection terms: the zonal advection of $u_R$ by the mean flow, and the meridional advection of the mean flow by $\upsilon_R$. Both terms have maxima in the subtropics as shown in Fig.~\ref{fig:tend_uZ}a for the initial time step. They have opposite signs, but the zonal advection term is more than twice as large and dominates the total effect. 

The contribution of the forced Rossby wave-mean flow interactions to the depth tendency
\begin{equation}
    N_h = \underbrace{-\frac{\gamma}{\cos\varphi} u_0\frac{\partial h_R}{\partial\lambda}-\gamma\upsilon_R\frac{\partial h_0}{\partial\varphi}}_\mathrm{advection} \underbrace{-\gamma h_0\nabla\cdot \mathbf{V}_R}_\mathrm{divergence}\label{eq:tend_z}
\end{equation}
is caused by advection of depth, and the Rossby wave divergence on the sphere, multiplied by the zonal mean depth perturbation. The two terms have opposite signs, but the amplitude of the divergence term is somewhat greater, making the total tendency $N_h$ negative throughout the tropics to 30$^\circ$N (Fig.~\ref{fig:tend_uZ}b). The advection term, which is small, would be zero if $u_0$ and $\upsilon_R$ obeyed the linear geostrophic balance. The magnitude of the divergence term is determined by the relatively large $h_0$ of the balanced background flow.

The Kelvin wave tendencies due to $N_u$, $N_\upsilon$ and $N_h$ are evaluated using the Hough transform, and the sum of these tendencies is $\hat{f}$ in Eq.~(\ref{eq:kw_prognostic_forced}). The largest contribution to the $u_K$ tendency in REF is caused by the zonal advection of $u_R$ (Fig. \ref{fig:tend_uZ}c). This is due to the large magnitude of $u_R$ around $20^\circ$N (Fig.~\ref{fig:tend_uZ}a), where the KW is still distinctly larger than zero. The meridional advection of zonal velocity has a weaker and opposite effect on the KWs, and the effect of the depth advection is even weaker (Fig.~\ref{fig:tend_uZ}c). The depth tendency due to the Rossby wave divergence is the second largest contributor to the KW tendencies. In agreement with the small meridional velocity of the KW, which is zero on the equatorial $\beta$-plane, the influence of the $\upsilon$ tendency on the KW is negligible.

\begin{figure}[t]
    \centering
    \includegraphics[width=
.7\textwidth]{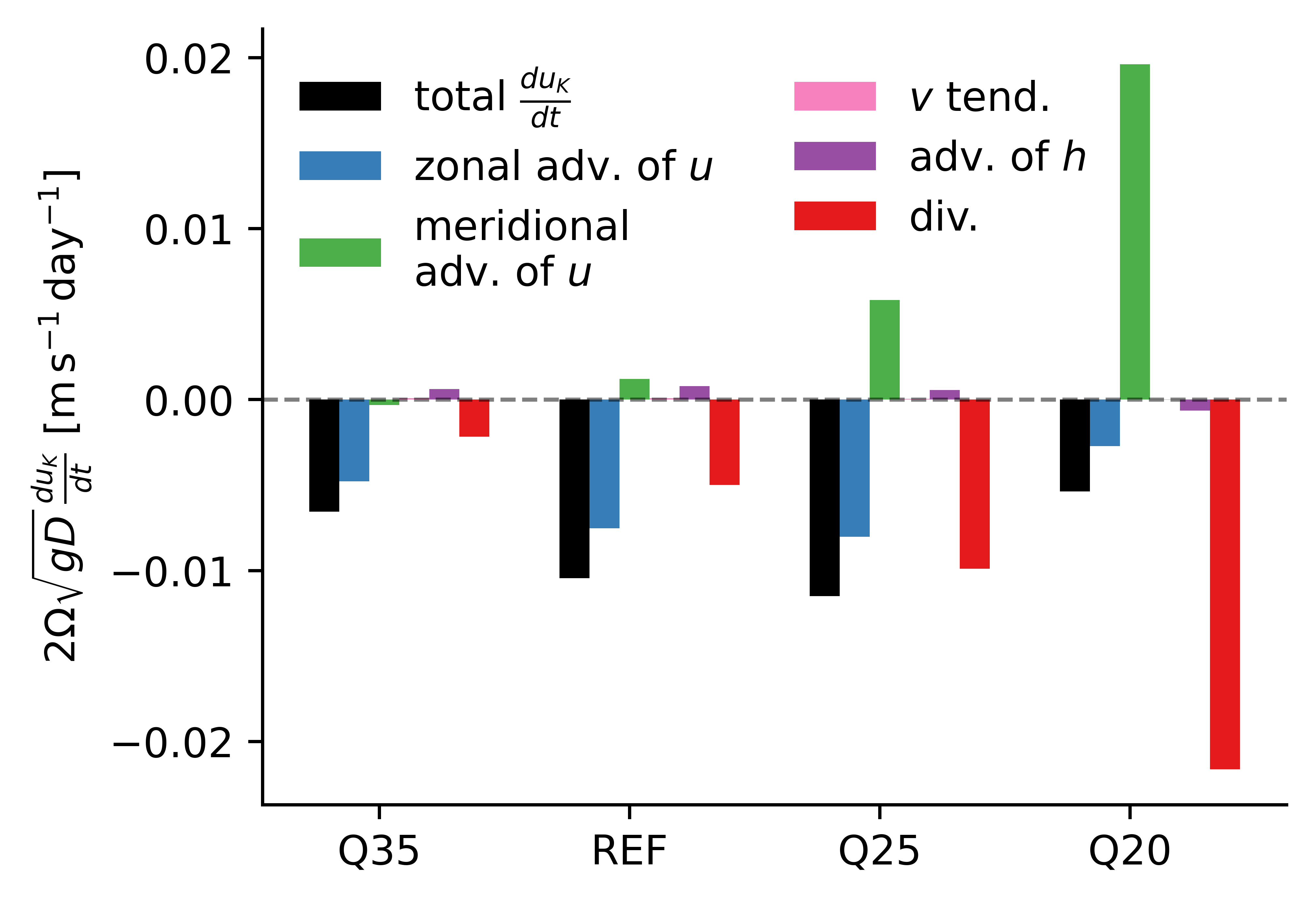}
    \caption{Zonal velocity tendencies of the $k=1$ Kelvin wave due to different terms of the forced Rossby wave-mean flow interactions at 0$^\circ$E, 1.4$^\circ$N for simulations in which the Rossby wave forcing is centered at different latitudes.}
    \label{fig:tend_compare}
\end{figure}

In the following, we investigate the relative importance of different terms contributing to the KW tendency in simulations with a poleward or equatorward-shifted forcing with respect to REF. The amplitudes of different terms in Q20, Q25 and Q35 are presented in Fig.~\ref{fig:tend_compare} along with REF. The comparison shows that the meridional advection term increases when the forcing is closer to the equator. This is caused by the stronger shear of the zonal background flow in that region, as well as the greater overlap of the Rossby wave-mean flow interactions with the meridional Kelvin wave structure. 
However, the Kelvin wave tendencies due to the meridional advection term are outweighed by the large KW tendencies due to the divergence term, which have opposite sign. The latter are stronger when the forcing is shifted equatorward, so that the Rossby waves are located in regions of larger $h_0$. Therefore, the KW tendencies due to the divergence term in Q20 and Q25 exceed the KW tendencies due to the zonal advection of $u_R$ (Fig.~\ref{fig:tend_compare}). 

For Q25, the zonal advection term has a larger magnitude than in REF (Fig.~\ref{fig:tend_compare}), because of the greater overlap of the Rossby wave-mean flow interactions with the meridional structure of the Kelvin waves. When the Rossby waves are located even closer to the equator (Q20), their zonal velocities are collocated with weaker $u_0$, which reduces the zonal advection term. Conversely, the Rossby waves are located in regions of stronger $u_0$ in Q35, but the resulting tendencies are further away from the equator so that their influence on the Kelvin waves is weaker than in REF.
The KW tendencies due to the depth advection and due to the meridional velocity tendencies have in all simulations a similar magnitude as in REF (Fig.~\ref{fig:tend_compare}). The total Kelvin wave tendencies are largest for Q25.

The presented sensitivity experiments demonstrate that the KW tendencies depend on the latitude of the forcing. Similarly, it can be expected to find sensitivity to the background flow and the prescribed mean depth which defines the meridional scale of the KW. Quantifying these sensitivities is beyond the scope of our study.
Taking a different approach, \citet{barpanda_role_2023} found that the meridional advection of the background vorticity (Sverdrup effect) caused eddy divergence. In the equatorial region, this eddy divergence mainly belonged to the Kelvin modes. The Sverdrup effect is not easily compared with the momentum advection in our study, as \citet{barpanda_role_2023} analyzed different terms of the linearized vorticity equation for a steady state including meridional velocity, while we study the transient KW growth in a background with zero meridional velocity.
However, the background meridional velocity in the setup of \citet{barpanda_role_2023} might have had a negligible effect, because the dominant Sverdrup effect only contains the Rossby wave meridional velocity, not that of the mean flow.

\section{Conclusions and Outlook}
\label{sect:conclusions}
We have investigated the KW excitation by interactions of the subtropical Rossby waves and the zonal mean flow on the sphere. Special aspects of the applied modeling framework are the balanced background jet and the external Rossby wave forcing. Such a forcing also affects the Kelvin mode modified with respect to the solution for the resting background state so that it contains a small contribution from the Rossby modes (appendix B). The modification arises from the effects of Rossby wave-mean flow interactions. However, we can not use the modified eigenmode to define the Kelvin wave, because it is ambiguous when the background flow varies, and because the modified eigenmodes do not form an orthogonal basis (appendix A). The KW definition as an eigenmode of the background state of rest is a meaningful way to analyze the real atmosphere with constantly varying background flows and instabilities.

As the KW is a prognostic variable of our model, the individual terms contributing to the KW energy growth in both physical and spectral space can be separated into wave-mean flow and wave-wave interactions. In spectral space, we have found that the wave-wave interactions are weak compared to the wave-mean flow interactions, because the mean flow is stronger than the waves. In physical space, the advection of the zonal Rossby wave velocity by the background flow causes part of the KW energy growth. The influence of the meridional advection of background zonal velocity by Rossby waves increases when the Rossby waves are collocated with stronger shear. The resulting KW tendencies are outweighed by the tendencies due to Rossby wave divergence in regions of large deviations from the mean depth. However, such a strong influence of the divergence term is associated with the formulation of the balanced background flow in the simplified model. The relative importance of the terms in the real atmosphere or three-dimensional (3D) models may be different.

Assuming stationary mean flow, the KW energy growth due to forcing with a prescribed frequency can be computed analytically. We showed a qualitative agreement between the temporal evolution of the KW energy in the resonance and non-resonance cases and the nonlinear simulations, especially for low-frequency forcing.  

Our simulations are characterized by Rossby and Froude numbers  smaller than one, and stable modes. 
For larger Rossby and Froude numbers and for stronger flows, instabilities might appear. 
For example, \citet{Wang.Mitchell_2014} discussed unstable Rossby-Kelvin modes for Rossby numbers around one and Froude numbers of about 1-3, in the context of equatorial superrotation found on other planets. 
Furthermore, Kelvin waves can be excited by barotropic instability for weak surface-to-pole temperature gradients and can contribute to the superrotation \citep{Polichtchouk.Cho_2016}.

While the KW excitation by convection is a well-studied process theoretically \citep[e.g.][]{Salby.Garcia_1987} and in data \citep[e.g.][]{Bergman.Salby_1994,Kiladisetal2009}, nonlinear aspects of KW dynamics have yet to be fully understood. The near non-dispersiveness and semi-geostrophy of the KW make the understanding of its variability and predictability only seemingly simpler compared to other large-scale equatorial modes.
The impacts of wave-mean flow interactions and resonance effects on Kelvin waves in the atmosphere are subject of further work including the relative roles of the wave-mean flow interactions in comparison with wave-wave interactions and convective forcing in reanalysis data and climate models. Idealized simulations with a 3D version of the TIGAR model, currently under development, might also provide an opportunity to research the coupling of the Madden-Julian Oscillation with subtropical Rossby wave dynamics \citep{Wedi.Smolarkiewicz_2010,Yano.Tribbia_2017}.

\clearpage
\acknowledgments
This study is funded by the German Science Foundation DFG, Grant No. 461186383, and by the Federal Ministry of Education and Research (BMBF) and the Free and Hanseatic City of Hamburg under the Excellence Strategy of the Federal Government and the Länder. We are very grateful to Rupert Klein and two anonymous reviewers for their insightful comments, and to our colleague Sándor István Mahó for TIGAR-related discussions and for reading the manuscript.

%
%
\datastatement
The output of the model simulations is available at https://doi.org/10.5281/zenodo.10673967.

%



\appendix[A]
\appendixtitle{Eigenmodes of the linearized equations around a non-resting background state}
Hough harmonics, which provide the basis of spectral expansion in the TIGAR model, are defined as 
eigensolutions of Laplace tidal equations, i.e.~equation \eqref{eq:SWM_physical} linearized around the state of no motion.  Since the linear operator $\mathbf{L}$ in Laplace tidal equation \eqref{LTE} is $L^2$ skew-adjoint, its non-zero eigenvalues are purely imaginary and the corresponding eigenfunctions (Hough harmonics) are mutually $L^2$-orthogonal. The $L^2$ inner product is
\begin{equation} \label{eq:scal}
 \langle \mathbf{X_1} \,, \mathbf{X_2} \rangle= \frac{1}{2\pi} \int_0^{2\pi} \int_{-\pi/2}^{\pi/2}\mathbf{X_1} \cdot\mathbf{X_2^*} \cos\varphi\; d\varphi\,d\lambda \, ,
\end{equation}
where $\mathbf{X_i} = (u_i,v_i,h_i)^T$ for $ i=1,2$.
Furthermore, \citet{Longuet-Higgins_1968} demonstrated that $\mathbf{L}$  possesses a purely point spectrum and thus, the set of Hough harmonics is $L^2$ complete. 

This remarkable analytical structure is destroyed when the nondimensional RSW equations \eqref{eq:SWM_physical} are linearized around a nontrivial zonal steady state $(u_0(\varphi)$,\,0,\, $h_0(\varphi))^\intercal$.  In that case the linearized equations read
\begin{equation} \label{eq:LTEmod}
    \frac{\partial}{\partial t} \mathbf{X} + \mathbf{L_b}\mathbf{X} = 0,
\end{equation}
with the modified linear operator 
\begin{equation}
     \mathbf{L_b}= \mathbf{L}+\mathbf{B} \,, 
\end{equation}
where $\mathbf{B}$ contains the background flow terms, 
\begin{equation}
    \mathbf{B} = \begin{pmatrix}
        \bar{\omega} \frac{\partial }{ \partial \lambda} & \left(\cos\varphi \frac{d\bar{\omega}}{d\varphi} - 2\bar{\omega}\sin\varphi \right) & 0\\
        2\bar{\omega}\sin\varphi & \bar{\omega} \frac{\partial}{\partial \lambda} & 0\\
        \frac{\gamma h_0}{\cos\varphi} \frac{\partial}{\partial \lambda} & \frac{\gamma}{ \cos \varphi}  \frac{ \partial }{ \partial \varphi}[h_0 \cos \varphi\,  (\cdot )] & \bar{\omega}\frac{\partial}{\partial \lambda}
    \end{pmatrix}
\end{equation}
 and $\bar{\omega}(\varphi) = \gamma u_0/\cos(\varphi)$.

Let $i\sigma$ be an eigenvalue of $ \mathbf{L_b}$. 
 We refer to $\mathrm{Re}(\sigma)$ as {\em modified frequency} and call the corresponding eigenfunction of $\mathbf{L_b}$ a {\em modified eigenmode}. It must be noted that the spectrum of $\mathbf{L_b}$ is not necessarily purely point. For instance, for certain background flows the continuous spectrum is not empty \citep{Kasahara_1980,Mitchell_2013}. 

Moreover, for an arbitrary non-trivial steady state background, the modified eigenmodes do not form an orthogonal $L^2$ basis. This has been mentioned by \citet{Dickinson.Williamson_1972} (for $h_0=0$) and other authors \citep[e.g.][]{teruya.etal_2024}, however, the proof of this statement seems to be missing in the literature.

\begin{proposition} The set of modified eigenmodes $\{ \mathbf{e_j} \} $ is not an orthogonal basis in $L^2$, unless $\bar \omega $ is constant and $h_0 = 0$. Furthermore, under the same assumptions, the set of $\{ \mathbf{P_k} \mathbf{e_j} \} $ , where $\mathbf{P_k}$ is the projector onto the fields with zonal wave number $k$ is not an orthogonal basis in $L^2([-\pi/2, \pi/2])$ with the scalar product 
\begin{equation} \label{eq:scal}
 \langle \mathbf{X_1} \,, \mathbf{X_2} \rangle_{\varphi}= \int_{-\pi/2}^{\pi/2}\mathbf{X_1} \cdot\mathbf{X_2^*} \cos\varphi\; d\varphi\, \, .
\end{equation}
\end{proposition}

\textbf{Proof}. From the assumption that the modified eigenmodes are an orthogonal basis, it follows that  on any finite-dimensional span of modified eigenmodes $\mathbf{L_b}$ can be expressed as $\mathbf{L_b} = \mathbf{U}\mathbf{T}\mathbf{U}^\ast$ with unitary operator $\mathbf{U}$ and diagonal operator $\mathbf{T}$. By completeness of the basis, this generalizes to the whole domain of $\mathbf{L_b}$. The adjoint of $\mathbf{L_b}$ is $\mathbf{U}\mathbf{T}^\ast\mathbf{U}^\ast$. Since 
\begin{equation}
\mathbf{L_b} \mathbf{L_b}^\ast = \mathbf{U}\mathbf{T} \mathbf{T}^\ast\mathbf{U}^\ast =\mathbf{U}\mathbf{T}^\ast \mathbf{T} \mathbf{U}^\ast = \mathbf{L_b}^\ast \mathbf{L_b} \,,
\end{equation}
it follows that $\mathbf{L_b}$ is normal. However, 
\begin{equation}
 \mathbf{L_b}^\ast\mathbf{L_b} - \mathbf{L_b}\mathbf{L_b}^\ast  =
\begin{pmatrix}
 \alpha + \frac{(1-h_1^2)\gamma^2}{\cos^2\varphi} \frac{\partial^2}{\partial\lambda^2}  & x_{12} & \gamma\beta h_0\frac{\partial}{\partial\varphi} + \cos\varphi \frac{d\bar{\omega}}{d\varphi} \gamma \frac{\partial}{\partial\varphi} \\
        x_{21} & x_{22} & x_{23}\\
        x_{31} & x_{32} & x_{33}
    \end{pmatrix} \,,
\end{equation}
where $\alpha= (2\bar \omega +1)\sin 2 \varphi \frac{\partial \bar \omega}{\partial \varphi} - (\cos \varphi \frac{\partial \bar \omega}{\partial \varphi})^2$, $\beta = \cos \varphi \frac{\partial \bar \omega}{\partial \varphi}- (2\bar \omega +1)\sin \varphi$,  $h_1 = h_0 +1$, and $x_{ij}$ denotes matrix elements which are not required for our conclusion. Each of the summands of $x_{11}$ in $\mathbf{L_b}^\ast\mathbf{L_b} - \mathbf{L_b}\mathbf{L_b}^\ast$ must be zero separately for $\mathbf{L_b}$ to be normal. From the second summand being zero, it follows that $h_0$ must be zero. With $h_0=0$, $x_{13}$ can only be zero if $\frac{d\bar{\omega}}{d\varphi}$ is zero. Thus, $\mathbf{L_b}$ is normal if and only if the background flow is absent, i.e.~$\mathbf{L_b} = \mathbf{L}$, or if $\bar{\omega}$ is constant and $\frac{d\bar{\omega}}{d\varphi} = h_0 = 0$, while the latter case of solid-body rotation without variations in the fluid depth is not a steady state. The statement that the projections $ \{ \mathbf{P_k} \mathbf{e_j}\}$ do not form an orthogonal basis 
follows by applying the same argument to the operator $\mathbf{L_b} \mathbf{P_k}$. 

Thus, for a non-trivial background flow, the modified eigenmodes are not mutually orthogonal, or they do not span the whole space, or both. For this reason, the modified eigenmodes, while dynamically important, are unsuitable as a basis of a forecast model or spectral expansion of atmospheric data. 

The orthogonality of modes in shear flows discussed by \citet{Held_1985} using the conservation of pseudomomentum relies on a different inner product. Furthermore, generalized Hough modes which take into account horizontal and vertical shear might not form a complete system, and they are for the lowest order orthogonal to their adjoints \citep{Ortland_2005}.

In order to compute the modified normal modes and corresponding eigenvalues, we use the FORTRAN program {\textrm{BGHough}}, which is part of TIGAR. This program implements the algorithm proposed in \citet{Kasahara_1980}, which we describe below.  

Substituting the ansatz 
\begin{equation}
    \mathbf{W}= (u,v,h)^\intercal = \mathbf{\widehat W } (\varphi) e^{i(k\lambda-\sigma t)}
\end{equation}
into the linarized RSW equation \eqref{eq:LTEmod}, yields 
\begin{equation}
    (\mathbf{L} - i\sigma \mathbf{I})\mathbf{\widehat W} + i\mathbf{\widehat B} \; \mathbf{\widehat W} = 0, \label{eq:linearized}
\end{equation}
where $\mathbf{I}$ is the identity matrix, and
\begin{equation}
    \mathbf{\widehat B} = \begin{pmatrix}
        \bar{\omega} k & i\left(2\bar{\omega}\sin\varphi - \cos\varphi \frac{d\bar{\omega}}{d\varphi}\right) & 0\\
        -2i\bar{\omega}\sin\varphi & \bar{\omega}k & 0\\
        \frac{k\gamma h_0}{\cos\varphi} & -\frac{i \gamma}{ \cos \varphi}  \frac{ \partial }{ \partial \varphi}[h_0 \cos \varphi\,  (\cdot )] & \bar{\omega}k
    \end{pmatrix} \,.
\end{equation}

To solve Eq.~(\ref{eq:linearized}), $\mathbf{\widehat W}$ is expanded in terms of Hough functions, so that 
\begin{equation}
    \mathbf{\widehat W} = \sum_r C_r \mathbf{\Theta}_r^{k} \label{eq:backgroundexpansion}
\end{equation}
where the sum is taken over all Hough modes with the zonal wavenumber $k$ available at the selected resolution. 

Substituting  Eq.~(\ref{eq:backgroundexpansion}) into Eq.~(\ref{eq:linearized}) yields an eigenvalue problem 
\begin{equation}\label{eq:eigM}
(\mathbf{M}- \sigma \mathbf{I})\mathbf{C} = 0    
\end{equation}
where $\mathbf{C}= (C_1 \hdots C_R)^\intercal$ and the matrix $\mathbf{M}$ is given by 
\begin{equation}
    \mathbf{M} = \begin{pmatrix}
        \nu_1^{k} + b_{11}^{k} & b_{12}^{k} & \hdots & b_{1R}^{k}\\
        b_{21}^{k} & \nu_2^{k} + b_{22}^{k} & \hdots & b_{2R}^{k}\\
        \vdots & \vdots & \, & \vdots\\ 
        b_{R1}^{k} & b_{R2}^{k} & \hdots & \nu_R^{k} +b_{RR}^{k}\\
    \end{pmatrix}
\end{equation}
with
\begin{equation}
    b_{r'r}^{k} = \int_{-1}^1 \mathbf{\widehat B} \mathbf{\Theta}_r^{k}\cdot \left(\mathbf{\Theta}_{r'}^{k}\right)^\ast\, .
\end{equation}
The eigenvalue problem \eqref{eq:eigM} is solved using DGEEV routine from LAPACK package and the resulting coefficients are normalized so that $\sum_{r}|C_{r}|^2 = 1$. 

We remark that if $\sigma_r$ is an eigenvalue of problem \eqref{eq:eigM}, then $i \sigma_r$ is the eigenvalue of the linearized RSW \eqref{eq:LTEmod}. Eigenvalues $i \sigma_r$, are not necessarily imaginary. The modified normal modes corresponding to $\mathrm{Im}(\sigma_r) > 0  $ are linearly unstable, while the ones corresponding to $\mathrm{Im}(\sigma_r) < 0  $ are exponentially stable, with growth and decay exponents, respectively, given by $\mathrm{Im}(\sigma_r)$. In all cases, the frequency of the modified mode is $\mathrm{Re}(\sigma_r)$. 

Classifying the modified modes is a non-trivial task since a modified mode can, in principle, project on all Hough harmonics with the same zonal wave number. \citet{Boyd_1978a} used the perturbation theory to define a modified Kelvin mode on the equatorial $\beta$-plane. \citet{Zhang.Webster_1989} classify the modified modes on the $\beta$-plane using an approximate dispersion relationship. This strategy is not currently feasible in spherical geometry as no such explicit approximate dispersion relationship is known.   

For the sphere,  there are at least two sensible classification strategies that are easy to implement. 
The first one is based on matching the spatial structures of modified modes and Hough harmonics. This defines the modified Kelvin mode as the one with the largest projection onto the "true" Kelvin wave (i.e. Kelvin mode for the state of no motion). The disadvantage of this approach is that for any fixed $n,l,k$, the modified frequencies of the modes with the largest projection on $\mathbf{H}^k_{n,l}$, 
$\mathbf{H}^{k\pm 1}_{n,l}$, and $\mathbf{H}^{k}_{n \pm 1,l}$ may significantly differ, resulting in non-smooth and non-monotone dispersion curves. 

We solve this problem by adopting a hybrid approach based on both the spatial structures  and the modified frequencies. First, a third of the modified modes with the largest "true" Rossby contribution, i.e. $\sum_r'\left|C_{r'}\right|^2$ with $r'$ running over all Rossby and MRG Hough modes, is classified as Rossby/MRG. The remaining modes are considered IG. Within the Rossby and IG categories, the classification is based on the modified frequency $\nu_{r, mod} = \mathrm {Re} (\sigma_r)$. The modified Rossby modes for a given $k$ are sorted according to $\nu_{r, mod}$ and the one with the lowest frequency is classified as modified MRG ($n=0$ modified mode), the mode with the second lowest freqeuncy is classified as $n=1$ and so on. Similarly, the modified IG modes are also sorted according to $\nu_{r,mod}$, and split into two bins at the median frequency. The bin containing the lowest frequencies corresponds to WIGs, while the other bin corresponds to EIGs. The modified Kelvin mode is then defined as the mode with the lowest frequency in the second bin ($n=0$ modified EIG). We note that throughout the procedure the frequency is treated as a signed quantity and "lowest" and "highest" should be interpreted accordingly.

\appendix[B]
\appendixtitle{Modified eigenmodes for the background flow in this study}
In the following, the eigenmodes with $k=1$ of the linearized equations with respect to the background flow in the present study (Fig.~\ref{fig:zf4}) are discussed. We identify the modified Kelvin mode as explained in appendix A. In this case it is not important, which of the alternative strategies is used for the task, as they both yield the same result. All of the modified Kelvin modes are stable and not decaying. 

\begin{figure}[t]
    \centering
    \includegraphics[width=.5\textwidth]{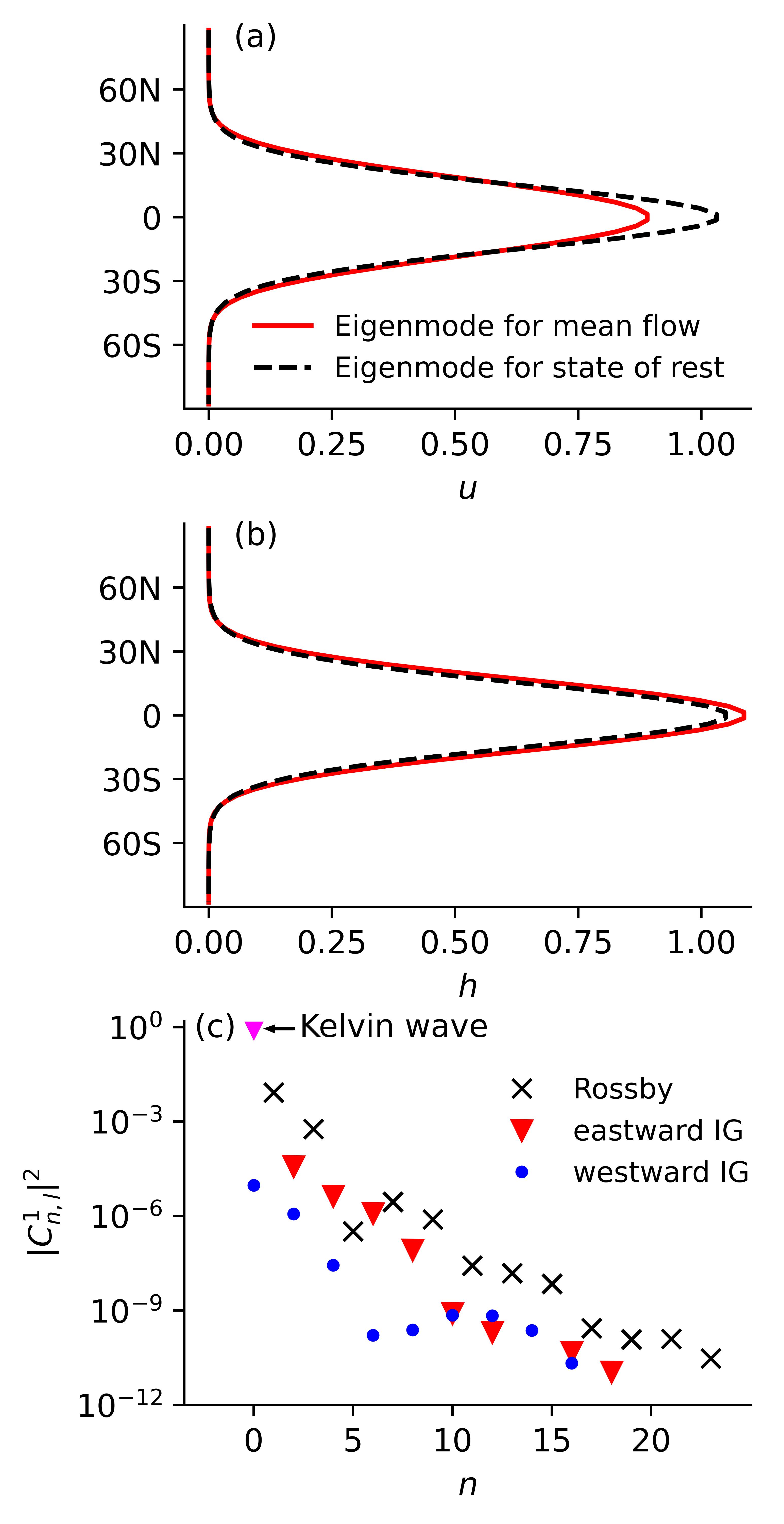}
    \caption{Meridional structure of (a) zonal velocity and (b) fluid depth of the modified eigenmode corresponding to the $k=1$ Kelvin mode for the background flow in Fig.~\ref{fig:zf4}. The Hough harmonic spectrum (c) contains only symmetric modes, which have even $n$ for IG modes and uneven $n$ for Rossby modes.}
    \label{fig:kw_bg}
\end{figure}

The modified Kelvin modes also have a similar spatial structure as Kelvin Hough modes for the corresponding zonal wave number (Fig.~\ref{fig:kw_bg}a,b). The differences in the meridional structure of the fluid depth are smaller than the differences in zonal velocity. Quantitatively, 99 \% of the energy of the modified $k=1$ Kelvin mode is carried by the Hough Kelvin wave ( $|C_{0,1}^1|^2 = 0.99$ ). The remaining 1 \% is the contribution of $n=1$ and $n=3$ Rossby modes  (Fig. \ref{fig:kw_bg}c). 

\begin{figure}[t]
    \centering
    \includegraphics[width=.5\textwidth]{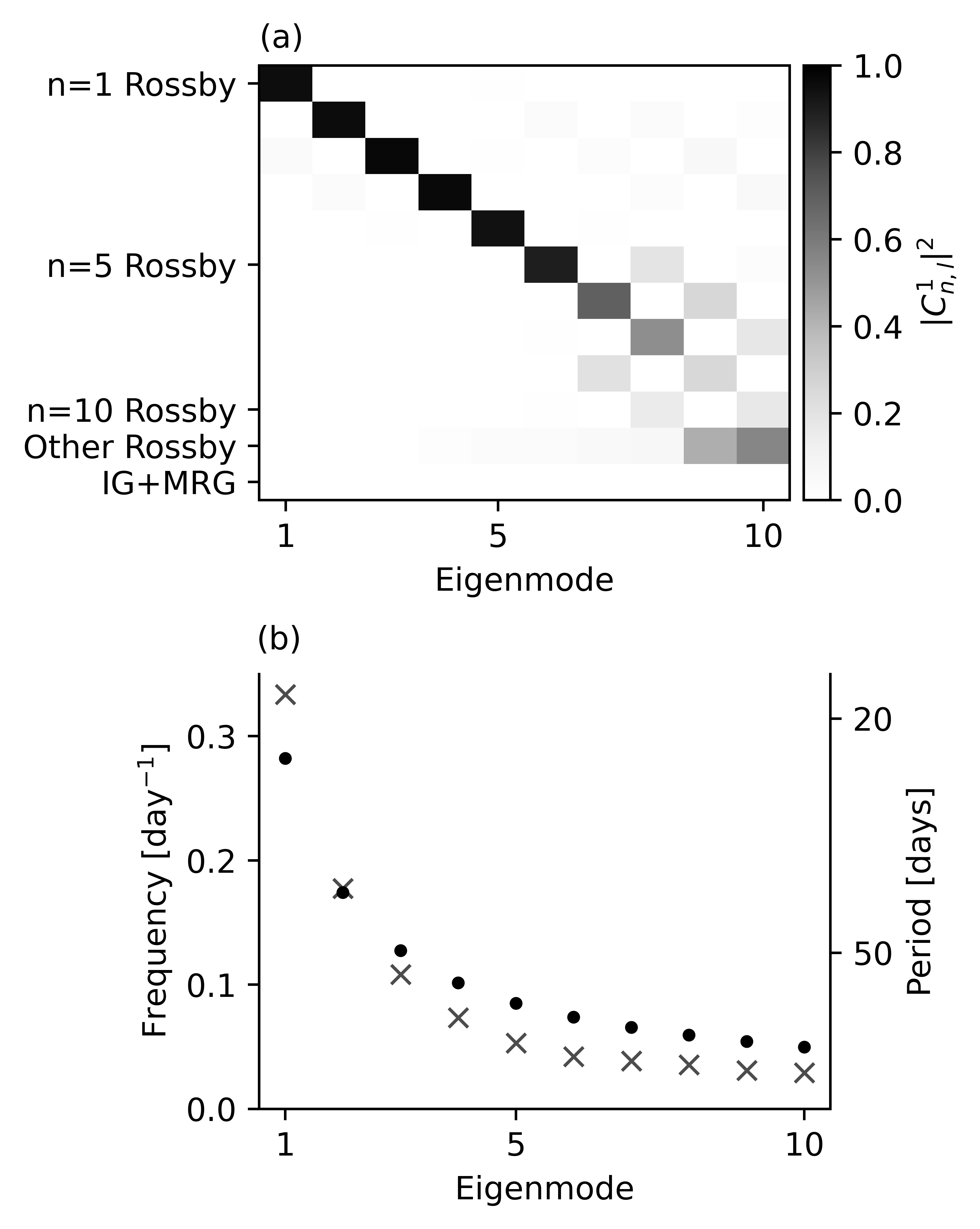}
    \caption{(a) Hough harmonic spectrum of the $k=1$ eigensolutions of the RSW equations, linearized around the background flow in Fig.~\ref{fig:zf4}, which correspond to the $n=1\hdots 10$ Rossby modes. (b) Crosses: Absolute frequencies of these modes as a function of $n$. Dots: frequencies for the state of rest.}
    \label{fig:frequencies_k1}
\end{figure}

The modified Rossby modes with $k=1$ and low $n$ also match well with the respective Hough harmonics (Fig. \ref{fig:frequencies_k1}a). This is in line with \citet{Kasahara_1980_corr}, and small changes of the lowest Rossby modes through shear flows have also been found by \citet{Boyd_1978a} and \citet{Mitchell_2013}. For higher $n$, several Rossby modes contribute to the modified structure, but the IG wave contribution is small for all shown modes (Fig. \ref{fig:frequencies_k1}a).

The frequencies of the lowest ten modified Rossby modes are shown in Fig.~\ref{fig:frequencies_k1}b. For small $n$, the frequencies with respect to the balanced background state in Fig.~\ref{fig:zf4} are higher than the frequencies for the state of rest because the increase of the absolute frequency induced by $h_0$ outweighs the decrease by the Doppler shift. The frequencies decrease with $n$, and the differences are mostly small. 

\clearpage



%



\bibliographystyle{ametsocV6}
\bibliography{references}

\end{document}